\magnification=\magstep1 
\font\bigbfont=cmbx10 scaled\magstep1
\font\bigifont=cmti10 scaled\magstep1
\font\bigrfont=cmr10 scaled\magstep1
\font\smallrfont=cmr6 scaled\magstep1
\font\smallitfont=cmti6 scaled\magstep1
\vsize = 23.5 truecm
\hsize = 15.5 truecm
\hoffset = .2truein
\baselineskip = 14 truept
\overfullrule = 0pt
\parskip = 3 truept
\def\fmtname{AmS-TeX}

\def\fmtversion{2.1}
\catcode`\@=11
\ifx\amstexloaded@\relax\catcode`\@=\active
   \else\let\amstexloaded@\relax\fi
\newlinechar=`\^^J
\def\W@{\immediate\write\sixt@@n}
\def\CR@{\W@{^^J\fmtname - Version \fmtversion^^J%
COPYRIGHT 1985, 1990, 1991 - AMERICAN MATHEMATICAL SOCIETY^^J%
Use of this macro package is not restricted provided^^J%
each use is acknowledged upon publication.^^J}}
\CR@ \everyjob{\CR@}
\message{Loading definitions for}
\message{misc utility macros,}
\toksdef\toks@@=2
\long\def\rightappend@#1\to#2{\toks@{\\{#1}}\toks@@
 =\expandafter{#2}\xdef#2{\the\toks@@\the\toks@}\toks@{}\toks@@{}}
\def\alloclist@{}
\newif\ifalloc@
\def\showallocations{{\def\\{\immediate\write\m@ne}\alloclist@}\alloc@true}
\def\alloc@#1#2#3#4#5{\global\advance\count1#1by\@ne
 \ch@ck#1#4#2\allocationnumber=\count1#1
 \global#3#5=\allocationnumber
 \edef\next@{\string#5=\string#2\the\allocationnumber}%
 \expandafter\rightappend@\next@\to\alloclist@}
\newcount\count@@
\newcount\count@@@
\def\FN@{\futurelet\next}
\def\DN@{\def\next@}
\def\DNii@{\def\nextii@}
\def\RIfM@{\relax\ifmmode}
\def\RIfMIfI@{\relax\ifmmode\ifinner}
\def\setboxz@h{\setbox\z@\hbox}
\def\wdz@{\wd\z@}
\def\boxz@{\box\z@}
\def\setbox@ne{\setbox\@ne}
\def\wd@ne{\wd\@ne}
\def\iterate{\body\expandafter\iterate\else\fi}
\def\err@#1{\errmessage{AmS-TeX error: #1}}
\newhelp\defaulthelp@{Sorry, I already gave what help I could...^^J
Maybe you should try asking a human?^^J
An error might have occurred before I noticed any problems.^^J
``If all else fails, read the instructions.''}
\def\Err@{\errhelp\defaulthelp@\err@}
\def\eat@#1{}
\def\in@#1#2{\def\in@@##1#1##2##3\in@@{\ifx\in@##2\in@false\else\in@true\fi}%
 \in@@#2#1\in@\in@@}
\newif\ifin@
\def\space@.{\futurelet\space@\relax}
\space@. %
\newhelp\athelp@
{Only certain combinations beginning with @ make sense to me.^^J
Perhaps you wanted \string\@\space for a printed @?^^J
I've ignored the character or group after @.}
{\catcode`\~=\active 
 \lccode`\~=`\@ \lowercase{\gdef~{\FN@\at@}}}
\def\at@{\let\next@\at@@
 \ifcat\noexpand\next a\else\ifcat\noexpand\next0\else
 \ifcat\noexpand\next\relax\else
   \let\next\at@@@\fi\fi\fi
 \next@}
\def\at@@#1{\expandafter
 \ifx\csname\space @\string#1\endcsname\relax
  \expandafter\at@@@ \else
  \csname\space @\string#1\expandafter\endcsname\fi}
\def\at@@@#1{\errhelp\athelp@ \err@{\Invalid@@ @}}
\def\atdef@#1{\expandafter\def\csname\space @\string#1\endcsname}
\newhelp\defahelp@{If you typed \string\define\space cs instead of
\string\define\string\cs\space^^J
I've substituted an inaccessible control sequence so that your^^J
definition will be completed without mixing me up too badly.^^J
If you typed \string\define{\string\cs} the inaccessible control sequence^^J
was defined to be \string\cs, and the rest of your^^J
definition appears as input.}
\newhelp\defbhelp@{I've ignored your definition, because it might^^J
conflict with other uses that are important to me.}
\def\define{\FN@\define@}
\def\define@{\ifcat\noexpand\next\relax
 \expandafter\define@@\else\errhelp\defahelp@                               
 \err@{\string\define\space must be followed by a control
 sequence}\expandafter\def\expandafter\nextii@\fi}                          
\def\undefined@@@@@@@@@@{}
\def\preloaded@@@@@@@@@@{}
\def\next@@@@@@@@@@{}
\def\define@@#1{\ifx#1\relax\errhelp\defbhelp@                              
 \err@{\string#1\space is already defined}\DN@{\DNii@}\else
 \expandafter\ifx\csname\expandafter\eat@\string                            
 #1@@@@@@@@@@\endcsname\undefined@@@@@@@@@@\errhelp\defbhelp@
 \err@{\string#1\space can't be defined}\DN@{\DNii@}\else
 \expandafter\ifx\csname\expandafter\eat@\string#1\endcsname\relax          
 \global\let#1\undefined\DN@{\def#1}\else\errhelp\defbhelp@
 \err@{\string#1\space is already defined}\DN@{\DNii@}\fi
 \fi\fi\next@}

\def\predefine#1#2{\let#1#2}
\def\undefine#1{\let#1\undefined}
\message{page layout,}
\newdimen\captionwidth@
\captionwidth@\hsize
\advance\captionwidth@-1.5in
\def\pagewidth#1{\hsize#1\relax
 \captionwidth@\hsize\advance\captionwidth@-1.5in}

\def\hcorrection#1{\advance\hoffset#1\relax}
\def\vcorrection#1{\advance\voffset#1\relax}
\message{accents/punctuation,}

\let\graveaccent\`
\let\acuteaccent\'
\let\tildeaccent\~
\let\hataccent\^
\let\underscore\_
\let\B\=
\let\D\.
\let\ic@\/
\def\/{\unskip\ic@}
\def\textfonti{\the\textfont\@ne}
\def\t#1#2{{\edef\next@{\the\font}\textfonti\accent"7F \next@#1#2}}
\def~{\unskip\nobreak\ \ignorespaces}
\def\.{.\spacefactor\@m}
\atdef@;{\leavevmode\null;}
\atdef@:{\leavevmode\null:}
\atdef@?{\leavevmode\null?}
\edef\@{\string @}
\def\relaxnext@{\let\next\relax}
\atdef@-{\relaxnext@\leavevmode
 \DN@{\ifx\next-\DN@-{\FN@\nextii@}\else
  \DN@{\leavevmode\hbox{-}}\fi\next@}%
 \DNii@{\ifx\next-\DN@-{\leavevmode\hbox{---}}\else
  \DN@{\leavevmode\hbox{--}}\fi\next@}%
 \FN@\next@}
\def\srdr@{\kern.16667em}
\def\drsr@{\kern.02778em}
\def\sldl@{\drsr@}
\def\dlsl@{\srdr@}
\atdef@"{\unskip\relaxnext@
 \DN@{\ifx\next\space@\DN@. {\FN@\nextii@}\else
  \DN@.{\FN@\nextii@}\fi\next@.}%
 \DNii@{\ifx\next`\DN@`{\FN@\nextiii@}\else
  \ifx\next\lq\DN@\lq{\FN@\nextiii@}\else
  \DN@####1{\FN@\nextiv@}\fi\fi\next@}%
 \def\nextiii@{\ifx\next`\DN@`{\sldl@``}\else\ifx\next\lq
  \DN@\lq{\sldl@``}\else\DN@{\dlsl@`}\fi\fi\next@}%
 \def\nextiv@{\ifx\next'\DN@'{\srdr@''}\else
  \ifx\next\rq\DN@\rq{\srdr@''}\else\DN@{\drsr@'}\fi\fi\next@}%
 \FN@\next@}

\def\textfontii{\the\textfont\tw@}
\def\lbrace@{\delimiter"4266308 }
\def\rbrace@{\delimiter"5267309 }
\def\{{\RIfM@\lbrace@\else{\textfontii f}\spacefactor\@m\fi}
\def\}{\RIfM@\rbrace@\else
 \let\@sf\empty\ifhmode\edef\@sf{\spacefactor\the\spacefactor}\fi
 {\textfontii g}\@sf\relax\fi}
\let\lbrace\{
\let\rbrace\}
\def\AmSTeX{{\textfontii A\kern-.1667em%
  \lower.5ex\hbox{M}\kern-.125emS}-\TeX}
\message{line and page breaks,}
\def\vmodeerr@#1{\Err@{\string#1\space not allowed between paragraphs}}
\def\mathmodeerr@#1{\Err@{\string#1\space not allowed in math mode}}
\def\linebreak{\RIfM@\mathmodeerr@\linebreak\else
 \ifhmode\unskip\unkern\break\else\vmodeerr@\linebreak\fi\fi}

\newskip\saveskip@
\def\allowlinebreak{\RIfM@\mathmodeerr@\allowlinebreak\else
 \ifhmode\saveskip@\lastskip\unskip
 \allowbreak\ifdim\saveskip@>\z@\hskip\saveskip@\fi
 \else\vmodeerr@\allowlinebreak\fi\fi}
\def\nolinebreak{\RIfM@\mathmodeerr@\nolinebreak\else
 \ifhmode\saveskip@\lastskip\unskip
 \nobreak\ifdim\saveskip@>\z@\hskip\saveskip@\fi
 \else\vmodeerr@\nolinebreak\fi\fi}
\def\newline{\relaxnext@
 \DN@{\RIfM@\expandafter\mathmodeerr@\expandafter\newline\else
  \ifhmode\ifx\next\par\else
  \expandafter\unskip\expandafter\null\expandafter\hfill\expandafter\break\fi
  \else
  \expandafter\vmodeerr@\expandafter\newline\fi\fi}%
 \FN@\next@}
\def\dmatherr@#1{\Err@{\string#1\space not allowed in display math mode}}
\def\nondmatherr@#1{\Err@{\string#1\space not allowed in non-display math
 mode}}
\def\onlydmatherr@#1{\Err@{\string#1\space allowed only in display math mode}}
\def\nonmatherr@#1{\Err@{\string#1\space allowed only in math mode}}
\def\mathbreak{\RIfMIfI@\break\else
 \dmatherr@\mathbreak\fi\else\nonmatherr@\mathbreak\fi}
\def\nomathbreak{\RIfMIfI@\nobreak\else
 \dmatherr@\nomathbreak\fi\else\nonmatherr@\nomathbreak\fi}
\def\allowmathbreak{\RIfMIfI@\allowbreak\else
 \dmatherr@\allowmathbreak\fi\else\nonmatherr@\allowmathbreak\fi}
\def\pagebreak{\RIfM@
 \ifinner\nondmatherr@\pagebreak\else\postdisplaypenalty-\@M\fi
 \else\ifvmode\removelastskip\break\else\vadjust{\break}\fi\fi}
\def\nopagebreak{\RIfM@
 \ifinner\nondmatherr@\nopagebreak\else\postdisplaypenalty\@M\fi
 \else\ifvmode\nobreak\else\vadjust{\nobreak}\fi\fi}
\def\nonvmodeerr@#1{\Err@{\string#1\space not allowed within a paragraph
 or in math}}
\def\vnonvmode@#1#2{\relaxnext@\DNii@{\ifx\next\par\DN@{#1}\else
 \DN@{#2}\fi\next@}%
 \ifvmode\DN@{#1}\else
 \DN@{\FN@\nextii@}\fi\next@}
\def\newpage{\vnonvmode@{\vfill\break}{\nonvmodeerr@\newpage}}
\def\smallpagebreak{\vnonvmode@\smallbreak{\nonvmodeerr@\smallpagebreak}}
\def\medpagebreak{\vnonvmode@\medbreak{\nonvmodeerr@\medpagebreak}}
\def\bigpagebreak{\vnonvmode@\bigbreak{\nonvmodeerr@\bigpagebreak}}
\def\NoBlackBoxes{\global\overfullrule\z@}
\def\BlackBoxes{\global\overfullrule5\p@}
\def\Invalid@#1{\def#1{\Err@{\Invalid@@\string#1}}}
\def\Invalid@@{Invalid use of }
\message{figures,}
\Invalid@\caption
\Invalid@\captionwidth
\newdimen\smallcaptionwidth@
\def\topspace{\mid@false\ins@}
\def\midspace{\mid@true\ins@}
\newif\ifmid@
\def\captionfont@{}
\def\ins@#1{\relaxnext@\allowbreak
 \smallcaptionwidth@\captionwidth@\gdef\thespace@{#1}%
 \DN@{\ifx\next\space@\DN@. {\FN@\nextii@}\else
  \DN@.{\FN@\nextii@}\fi\next@.}%
 \DNii@{\ifx\next\caption\DN@\caption{\FN@\nextiii@}%
  \else\let\next@\nextiv@\fi\next@}%
 \def\nextiv@{\vnonvmode@
  {\ifmid@\expandafter\midinsert\else\expandafter\topinsert\fi
   \vbox to\thespace@{}\endinsert}
  {\ifmid@\nonvmodeerr@\midspace\else\nonvmodeerr@\topspace\fi}}%
 \def\nextiii@{\ifx\next\captionwidth\expandafter\nextv@
  \else\expandafter\nextvi@\fi}%
 \def\nextv@\captionwidth##1##2{\smallcaptionwidth@##1\relax\nextvi@{##2}}%
 \def\nextvi@##1{\def\thecaption@{\captionfont@##1}%
  \DN@{\ifx\next\space@\DN@. {\FN@\nextvii@}\else
   \DN@.{\FN@\nextvii@}\fi\next@.}%
  \FN@\next@}%
 \def\nextvii@{\vnonvmode@
  {\ifmid@\expandafter\midinsert\else
  \expandafter\topinsert\fi\vbox to\thespace@{}\nobreak\smallskip
  \setboxz@h{\noindent\ignorespaces\thecaption@\unskip}%
  \ifdim\wdz@>\smallcaptionwidth@\centerline{\vbox{\hsize\smallcaptionwidth@
   \noindent\ignorespaces\thecaption@\unskip}}%
  \else\centerline{\boxz@}\fi\endinsert}
  {\ifmid@\nonvmodeerr@\midspace
  \else\nonvmodeerr@\topspace\fi}}%
 \FN@\next@}
\message{comments,}
\def\newcodes@{\catcode`\\12\catcode`\{12\catcode`\}12\catcode`\#12%
 \catcode`\%12\relax}
\def\oldcodes@{\catcode`\\0\catcode`\{1\catcode`\}2\catcode`\#6%
 \catcode`\%14\relax}
\def\comment{\newcodes@\endlinechar=10 \comment@}
{\lccode`\0=`\\
\lowercase{\gdef\comment@#1^^J{\comment@@#10endcomment\comment@@@}%
\gdef\comment@@#10endcomment{\FN@\comment@@@}%
\gdef\comment@@@#1\comment@@@{\ifx\next\comment@@@\let\next\comment@
 \else\def\next{\oldcodes@\endlinechar=`\^^M\relax}%
 \fi\next}}}
\def\pr@m@s{\ifx'\next\DN@##1{\prim@s}\else\let\next@\egroup\fi\next@}
\def\prime{{\null\prime@\null}}
\mathchardef\prime@="0230
\let\dsize\displaystyle

\let\ssize\scriptstyle

\message{math spacing,}
\def\,{\RIfM@\mskip\thinmuskip\relax\else\kern.16667em\fi}
\def\!{\RIfM@\mskip-\thinmuskip\relax\else\kern-.16667em\fi}
\let\thinspace\,
\let\negthinspace\!
\def\medspace{\RIfM@\mskip\medmuskip\relax\else\kern.222222em\fi}
\def\negmedspace{\RIfM@\mskip-\medmuskip\relax\else\kern-.222222em\fi}
\def\thickspace{\RIfM@\mskip\thickmuskip\relax\else\kern.27777em\fi}
\let\;\thickspace
\def\negthickspace{\RIfM@\mskip-\thickmuskip\relax\else
 \kern-.27777em\fi}
\atdef@,{\RIfM@\mskip.1\thinmuskip\else\leavevmode\null,\fi}
\atdef@!{\RIfM@\mskip-.1\thinmuskip\else\leavevmode\null!\fi}
\atdef@.{\RIfM@&&\else\leavevmode.\spacefactor3000 \fi}
\def\and{\DOTSB\;\mathchar"3026 \;}

\message{fractions,}
\def\frac#1#2{{#1\over#2}}

\newdimen\ex@
\ex@.2326ex
\Invalid@\thickness
\def\thickfrac{\relaxnext@
 \DN@{\ifx\next\thickness\let\next@\nextii@\else
 \DN@{\nextii@\thickness1}\fi\next@}%
 \DNii@\thickness##1##2##3{{##2\above##1\ex@##3}}%
 \FN@\next@}

\def\thickfracwithdelims#1#2{\relaxnext@\def\ldelim@{#1}\def\rdelim@{#2}%
 \DN@{\ifx\next\thickness\let\next@\nextii@\else
 \DN@{\nextii@\thickness1}\fi\next@}%
 \DNii@\thickness##1##2##3{{##2\abovewithdelims
 \ldelim@\rdelim@##1\ex@##3}}%
 \FN@\next@}

\def\:{\nobreak\hskip.1111em\mathpunct{}\nonscript\mkern-\thinmuskip{:}\hskip
 .3333emplus.0555em\relax}
\def\snug{\unskip\kern-\mathsurround}
\message{smash commands,}
\def\topsmash{\top@true\bot@false\smash@}
\def\botsmash{\top@false\bot@true\smash@}
\newif\iftop@
\newif\ifbot@
\def\smash{\top@true\bot@true\smash@}
\def\smash@{\RIfM@\expandafter\mathpalette\expandafter\mathsm@sh\else
 \expandafter\makesm@sh\fi}
\def\finsm@sh{\iftop@\ht\z@\z@\fi\ifbot@\dp\z@\z@\fi\leavevmode\boxz@}
\message{large operator symbols,}
\def\LimitsOnSums{\global\let\slimits@\displaylimits}
\def\NoLimitsOnSums{\global\let\slimits@\nolimits}
\LimitsOnSums
\mathchardef\coprod@="1360       \def\coprod{\DOTSB\coprod@\slimits@}
\mathchardef\bigvee@="1357       \def\bigvee{\DOTSB\bigvee@\slimits@}
\mathchardef\bigwedge@="1356     \def\bigwedge{\DOTSB\bigwedge@\slimits@}
\mathchardef\biguplus@="1355     \def\biguplus{\DOTSB\biguplus@\slimits@}
\mathchardef\bigcap@="1354       \def\bigcap{\DOTSB\bigcap@\slimits@}
\mathchardef\bigcup@="1353       \def\bigcup{\DOTSB\bigcup@\slimits@}
\mathchardef\prod@="1351         \def\prod{\DOTSB\prod@\slimits@}
\mathchardef\sum@="1350          \def\sum{\DOTSB\sum@\slimits@}
\mathchardef\bigotimes@="134E    \def\bigotimes{\DOTSB\bigotimes@\slimits@}
\mathchardef\bigoplus@="134C     \def\bigoplus{\DOTSB\bigoplus@\slimits@}
\mathchardef\bigodot@="134A      \def\bigodot{\DOTSB\bigodot@\slimits@}
\mathchardef\bigsqcup@="1346     \def\bigsqcup{\DOTSB\bigsqcup@\slimits@}
\message{integrals,}
\def\LimitsOnInts{\global\let\ilimits@\displaylimits}
\def\NoLimitsOnInts{\global\let\ilimits@\nolimits}
\NoLimitsOnInts
\def\int{\DOTSI\intop\ilimits@}
\def\oint{\DOTSI\ointop\ilimits@}
\def\intic@{\mathchoice{\hskip.5em}{\hskip.4em}{\hskip.4em}{\hskip.4em}}
\def\negintic@{\mathchoice
 {\hskip-.5em}{\hskip-.4em}{\hskip-.4em}{\hskip-.4em}}
\def\intkern@{\mathchoice{\!\!\!}{\!\!}{\!\!}{\!\!}}
\def\intdots@{\mathchoice{\plaincdots@}
 {{\cdotp}\mkern1.5mu{\cdotp}\mkern1.5mu{\cdotp}}
 {{\cdotp}\mkern1mu{\cdotp}\mkern1mu{\cdotp}}
 {{\cdotp}\mkern1mu{\cdotp}\mkern1mu{\cdotp}}}
\newcount\intno@
\def\iint{\DOTSI\intno@\tw@\FN@\ints@}
\def\iiint{\DOTSI\intno@\thr@@\FN@\ints@}
\def\iiiint{\DOTSI\intno@4 \FN@\ints@}
\def\idotsint{\DOTSI\intno@\z@\FN@\ints@}
\def\ints@{\findlimits@\ints@@}
\newif\iflimtoken@
\newif\iflimits@
\def\findlimits@{\limtoken@true\ifx\next\limits\limits@true
 \else\ifx\next\nolimits\limits@false\else
 \limtoken@false\ifx\ilimits@\nolimits\limits@false\else
 \ifinner\limits@false\else\limits@true\fi\fi\fi\fi}
\def\multint@{\int\ifnum\intno@=\z@\intdots@                                
 \else\intkern@\fi                                                          
 \ifnum\intno@>\tw@\int\intkern@\fi                                         
 \ifnum\intno@>\thr@@\int\intkern@\fi                                       
 \int}                                                                      
\def\multintlimits@{\intop\ifnum\intno@=\z@\intdots@\else\intkern@\fi
 \ifnum\intno@>\tw@\intop\intkern@\fi
 \ifnum\intno@>\thr@@\intop\intkern@\fi\intop}
\def\ints@@{\iflimtoken@                                                    
 \def\ints@@@{\iflimits@\negintic@\mathop{\intic@\multintlimits@}\limits    
  \else\multint@\nolimits\fi                                                
  \eat@}                                                                    
 \else                                                                      
 \def\ints@@@{\iflimits@\negintic@
  \mathop{\intic@\multintlimits@}\limits\else
  \multint@\nolimits\fi}\fi\ints@@@}
\def\LimitsOnNames{\global\let\nlimits@\displaylimits}
\def\NoLimitsOnNames{\global\let\nlimits@\nolimits@}
\LimitsOnNames
\def\nolimits@{\relaxnext@
 \DN@{\ifx\next\limits\DN@\limits{\nolimits}\else
  \let\next@\nolimits\fi\next@}%
 \FN@\next@}
\message{operator names,}
\def\newmcodes@{\mathcode`\'"27\mathcode`\*"2A\mathcode`\."613A%
 \mathcode`\-"2D\mathcode`\/"2F\mathcode`\:"603A }
\def\operatorname#1{\mathop{\newmcodes@\kern\z@\fam\z@#1}\nolimits@}
\def\operatornamewithlimits#1{\mathop{\newmcodes@\kern\z@\fam\z@#1}\nlimits@}
\def\qopname@#1{\mathop{\fam\z@#1}\nolimits@}
\def\qopnamewl@#1{\mathop{\fam\z@#1}\nlimits@}
\def\arccos{\qopname@{arccos}}
\def\arcsin{\qopname@{arcsin}}
\def\arctan{\qopname@{arctan}}
\def\arg{\qopname@{arg}}
\def\cos{\qopname@{cos}}
\def\cosh{\qopname@{cosh}}
\def\cot{\qopname@{cot}}
\def\coth{\qopname@{coth}}
\def\csc{\qopname@{csc}}
\def\deg{\qopname@{deg}}
\def\det{\qopnamewl@{det}}
\def\dim{\qopname@{dim}}
\def\exp{\qopname@{exp}}
\def\gcd{\qopnamewl@{gcd}}
\def\hom{\qopname@{hom}}
\def\inf{\qopnamewl@{inf}}
\def\injlim{\qopnamewl@{inj\,lim}}
\def\ker{\qopname@{ker}}
\def\lg{\qopname@{lg}}
\def\lim{\qopnamewl@{lim}}
\def\liminf{\qopnamewl@{lim\,inf}}
\def\limsup{\qopnamewl@{lim\,sup}}
\def\ln{\qopname@{ln}}
\def\log{\qopname@{log}}
\def\max{\qopnamewl@{max}}
\def\min{\qopnamewl@{min}}
\def\Pr{\qopnamewl@{Pr}}
\def\projlim{\qopnamewl@{proj\,lim}}
\def\sec{\qopname@{sec}}
\def\sin{\qopname@{sin}}
\def\sinh{\qopname@{sinh}}
\def\sup{\qopnamewl@{sup}}
\def\tan{\qopname@{tan}}
\def\tanh{\qopname@{tanh}}
\def\varinjlim{\mathop{\vtop{\ialign{##\crcr
 \hfil\rm lim\hfil\crcr\noalign{\nointerlineskip}\rightarrowfill\crcr
 \noalign{\nointerlineskip\kern-\ex@}\crcr}}}}
\def\varprojlim{\mathop{\vtop{\ialign{##\crcr
 \hfil\rm lim\hfil\crcr\noalign{\nointerlineskip}\leftarrowfill\crcr
 \noalign{\nointerlineskip\kern-\ex@}\crcr}}}}
\def\varliminf{\mathop{\underline{\vrule height\z@ depth.2exwidth\z@
 \hbox{\rm lim}}}}

\newdimen\buffer@
\buffer@\fontdimen13 \tenex
\newdimen\buffer
\buffer\buffer@

\def\ResetBuffer{\fontdimen13 \tenex\buffer@\global\buffer\buffer@}
\def\shave#1{\mathop{\hbox{$\m@th\fontdimen13 \tenex\z@                     
 \displaystyle{#1}$}}\fontdimen13 \tenex\buffer}

\message{multilevel sub/superscripts,}
\Invalid@\\
\def\Let@{\relax\iffalse{\fi\let\\=\cr\iffalse}\fi}
\Invalid@\vspace
\def\vspace@{\def\vspace##1{\crcr\noalign{\vskip##1\relax}}}
\def\multilimits@{\bgroup\vspace@\Let@
 \baselineskip\fontdimen10 \scriptfont\tw@
 \advance\baselineskip\fontdimen12 \scriptfont\tw@
 \lineskip\thr@@\fontdimen8 \scriptfont\thr@@
 \lineskiplimit\lineskip
 \vbox\bgroup\ialign\bgroup\hfil$\m@th\scriptstyle{##}$\hfil\crcr}
\def\Sb{_\multilimits@}
\def\endSb{\crcr\egroup\egroup\egroup}
\def\Sp{^\multilimits@}

\def\spreadlines#1{\RIfMIfI@\onlydmatherr@\spreadlines\else
 \openup#1\relax\fi\else\onlydmatherr@\spreadlines\fi}
\def\Mathstrut@{\copy\Mathstrutbox@}
\newbox\Mathstrutbox@
\setbox\Mathstrutbox@\null
\setboxz@h{$\m@th($}
\ht\Mathstrutbox@\ht\z@
\dp\Mathstrutbox@\dp\z@
\message{matrices,}
\newdimen\spreadmlines@
\def\spreadmatrixlines#1{\RIfMIfI@
 \onlydmatherr@\spreadmatrixlines\else
 \spreadmlines@#1\relax\fi\else\onlydmatherr@\spreadmatrixlines\fi}
\def\format{\crcr\egroup\iffalse{\fi\ifnum`}=0 \fi\format@}
\newtoks\hashtoks@
\hashtoks@{#}
\def\format@#1\\{\def\preamble@{#1}%
 \def\l{$\m@th\the\hashtoks@$\hfil}%
 \def\c{\hfil$\m@th\the\hashtoks@$\hfil}%
 \def\r{\hfil$\m@th\the\hashtoks@$}%
 \edef\preamble@@{\preamble@}\ifnum`{=0 \fi\iffalse}\fi
 \ialign\bgroup\span\preamble@@\crcr}
\def\smallmatrix{\null\,\vcenter\bgroup\vspace@\Let@
 \baselineskip9\ex@\lineskip\ex@
 \ialign\bgroup\hfil$\m@th\scriptstyle{##}$\hfil&&\thickspace\hfil
 $\m@th\scriptstyle{##}$\hfil\crcr}
\def\endsmallmatrix{\crcr\egroup\egroup\,}

\newmuskip\dotsspace@
\dotsspace@1.5mu
\def\strip@#1 {#1}
\def\spacehdots#1\for#2{\multispan{#2}\xleaders
 \hbox{$\m@th\mkern\strip@#1 \dotsspace@.\mkern\strip@#1 \dotsspace@$}\hfill}
\def\hdotsfor#1{\spacehdots\@ne\for{#1}}
\def\multispan@#1{\omit\mscount#1\unskip\loop\ifnum\mscount>\@ne\sp@n\repeat}
\def\spaceinnerhdots#1\for#2\after#3{\multispan@{\strip@#2 }#3\xleaders
 \hbox{$\m@th\mkern\strip@#1 \dotsspace@.\mkern\strip@#1 \dotsspace@$}\hfill}
\def\innerhdotsfor#1\after#2{\spaceinnerhdots\@ne\for#1\after{#2}}
\def\endcases{\endmatrix\right.\egroup}
\message{multiline displays,}
\newif\ifinany@
\newif\ifinalign@
\newif\ifingather@
\def\strut@{\copy\strutbox@}
\newbox\strutbox@
\setbox\strutbox@\hbox{\vrule height8\p@ depth3\p@ width\z@}
\def\topaligned{\null\,\vtop\aligned@}
\def\botaligned{\null\,\vbox\aligned@}
\def\aligned{\null\,\vcenter\aligned@}
\def\aligned@{\bgroup\vspace@\Let@
 \ifinany@\else\openup\jot\fi\ialign
 \bgroup\hfil\strut@$\m@th\displaystyle{##}$&
 $\m@th\displaystyle{{}##}$\hfil\crcr}
\def\endaligned{\crcr\egroup\egroup}

\def\alignedat#1{\null\,\vcenter\bgroup\doat@{#1}\vspace@\Let@
 \ifinany@\else\openup\jot\fi\ialign\bgroup\span\preamble@@\crcr}
\newcount\atcount@
\def\doat@#1{\toks@{\hfil\strut@$\m@th
 \displaystyle{\the\hashtoks@}$&$\m@th\displaystyle
 {{}\the\hashtoks@}$\hfil}
 \atcount@#1\relax\advance\atcount@\m@ne                                    
 \loop\ifnum\atcount@>\z@\toks@=\expandafter{\the\toks@&\hfil$\m@th
 \displaystyle{\the\hashtoks@}$&$\m@th
 \displaystyle{{}\the\hashtoks@}$\hfil}\advance
  \atcount@\m@ne\repeat                                                     
 \xdef\preamble@{\the\toks@}\xdef\preamble@@{\preamble@}}

\def\gathered{\null\,\vcenter\bgroup\vspace@\Let@
 \ifinany@\else\openup\jot\fi\ialign
 \bgroup\hfil\strut@$\m@th\displaystyle{##}$\hfil\crcr}
\def\endgathered{\crcr\egroup\egroup}
\newif\iftagsleft@
\def\TagsOnLeft{\global\tagsleft@true}
\def\TagsOnRight{\global\tagsleft@false}
\TagsOnLeft
\newif\ifmathtags@
\def\TagsAsMath{\global\mathtags@true}
\def\TagsAsText{\global\mathtags@false}
\TagsAsText
\def\tagform@#1{\hbox{\rm(\ignorespaces#1\unskip)}}
\def\thetag{\leavevmode\tagform@}
\def\tag#1$${\iftagsleft@\leqno\else\eqno\fi                                
 \maketag@#1\maketag@                                                       
 $$}                                                                        
\def\maketag@{\FN@\maketag@@}
\def\maketag@@{\ifx\next"\expandafter\maketag@@@\else\expandafter\maketag@@@@
 \fi}
\def\maketag@@@"#1"#2\maketag@{\hbox{\rm#1}}                                
\def\maketag@@@@#1\maketag@{\ifmathtags@\tagform@{$\m@th#1$}\else
 \tagform@{#1}\fi}
\interdisplaylinepenalty\@M
\def\allowdisplaybreaks{\RIfMIfI@
 \onlydmatherr@\allowdisplaybreaks\else
 \interdisplaylinepenalty\z@\fi\else\onlydmatherr@\allowdisplaybreaks\fi}
\Invalid@\allowdisplaybreak
\Invalid@\displaybreak
\Invalid@\intertext
\def\allowdisplaybreak@{\def\allowdisplaybreak{\crcr\noalign{\allowbreak}}}
\def\displaybreak@{\def\displaybreak{\crcr\noalign{\break}}}
\def\intertext@{\def\intertext##1{\crcr\noalign{%
 \penalty\postdisplaypenalty \vskip\belowdisplayskip
 \vbox{\normalbaselines\noindent##1}%
 \penalty\predisplaypenalty \vskip\abovedisplayskip}}}
\newskip\centering@
\centering@\z@ plus\@m\p@
\def\align{\relax\ifingather@\DN@{\csname align (in
  \string\gather)\endcsname}\else
 \ifmmode\ifinner\DN@{\onlydmatherr@\align}\else
  \let\next@\align@\fi
 \else\DN@{\onlydmatherr@\align}\fi\fi\next@}
\newhelp\andhelp@
{An extra & here is so disastrous that you should probably exit^^J
and fix things up.}
\newif\iftag@
\newcount\and@
\def\align@{\inalign@true\inany@true
 \vspace@\allowdisplaybreak@\displaybreak@\intertext@
 \def\tag{\global\tag@true\ifnum\and@=\z@\DN@{&&}\else
          \DN@{&}\fi\next@}%
 \iftagsleft@\DN@{\csname align \endcsname}\else
  \DN@{\csname align \space\endcsname}\fi\next@}
\def\Tag@{\iftag@\else\errhelp\andhelp@\err@{Extra & on this line}\fi}
\newdimen\lwidth@
\newdimen\rwidth@
\newdimen\maxlwidth@
\newdimen\maxrwidth@
\newdimen\totwidth@
\def\measure@#1\endalign{\lwidth@\z@\rwidth@\z@\maxlwidth@\z@\maxrwidth@\z@
 \global\and@\z@                                                            
 \setbox@ne\vbox                                                            
  {\everycr{\noalign{\global\tag@false\global\and@\z@}}\Let@                
  \halign{\setboxz@h{$\m@th\displaystyle{\@lign##}$}
   \global\lwidth@\wdz@                                                     
   \ifdim\lwidth@>\maxlwidth@\global\maxlwidth@\lwidth@\fi                  
   \global\advance\and@\@ne                                                 
   &\setboxz@h{$\m@th\displaystyle{{}\@lign##}$}\global\rwidth@\wdz@        
   \ifdim\rwidth@>\maxrwidth@\global\maxrwidth@\rwidth@\fi                  
   \global\advance\and@\@ne                                                
   &\Tag@
   \eat@{##}\crcr#1\crcr}}
 \totwidth@\maxlwidth@\advance\totwidth@\maxrwidth@}                       
\def\displ@y@{\global\dt@ptrue\openup\jot
 \everycr{\noalign{\global\tag@false\global\and@\z@\ifdt@p\global\dt@pfalse
 \vskip-\lineskiplimit\vskip\normallineskiplimit\else
 \penalty\interdisplaylinepenalty\fi}}}
\def\black@#1{\noalign{\ifdim#1>\displaywidth
 \dimen@\prevdepth\nointerlineskip                                          
 \vskip-\ht\strutbox@\vskip-\dp\strutbox@                                   
 \vbox{\noindent\hbox to#1{\strut@\hfill}}
 \prevdepth\dimen@                                                          
 \fi}}
\expandafter\def\csname align \space\endcsname#1\endalign
 {\measure@#1\endalign\global\and@\z@                                       
 \ifingather@\everycr{\noalign{\global\and@\z@}}\else\displ@y@\fi           
 \Let@\tabskip\centering@                                                   
 \halign to\displaywidth
  {\hfil\strut@\setboxz@h{$\m@th\displaystyle{\@lign##}$}
  \global\lwidth@\wdz@\boxz@\global\advance\and@\@ne                        
  \tabskip\z@skip                                                           
  &\setboxz@h{$\m@th\displaystyle{{}\@lign##}$}
  \global\rwidth@\wdz@\boxz@\hfill\global\advance\and@\@ne                  
  \tabskip\centering@                                                       
  &\setboxz@h{\@lign\strut@\maketag@##\maketag@}
  \dimen@\displaywidth\advance\dimen@-\totwidth@
  \divide\dimen@\tw@\advance\dimen@\maxrwidth@\advance\dimen@-\rwidth@     
  \ifdim\dimen@<\tw@\wdz@\llap{\vtop{\normalbaselines\null\boxz@}}
  \else\llap{\boxz@}\fi                                                    
  \tabskip\z@skip                                                          
  \crcr#1\crcr                                                             
  \black@\totwidth@}}                                                      
\newdimen\lineht@
\expandafter\def\csname align \endcsname#1\endalign{\measure@#1\endalign
 \global\and@\z@
 \ifdim\totwidth@>\displaywidth\let\displaywidth@\totwidth@\else
  \let\displaywidth@\displaywidth\fi                                        
 \ifingather@\everycr{\noalign{\global\and@\z@}}\else\displ@y@\fi
 \Let@\tabskip\centering@\halign to\displaywidth
  {\hfil\strut@\setboxz@h{$\m@th\displaystyle{\@lign##}$}%
  \global\lwidth@\wdz@\global\lineht@\ht\z@                                 
  \boxz@\global\advance\and@\@ne
  \tabskip\z@skip&\setboxz@h{$\m@th\displaystyle{{}\@lign##}$}%
  \global\rwidth@\wdz@\ifdim\ht\z@>\lineht@\global\lineht@\ht\z@\fi         
  \boxz@\hfil\global\advance\and@\@ne
  \tabskip\centering@&\kern-\displaywidth@                                  
  \setboxz@h{\@lign\strut@\maketag@##\maketag@}%
  \dimen@\displaywidth\advance\dimen@-\totwidth@
  \divide\dimen@\tw@\advance\dimen@\maxlwidth@\advance\dimen@-\lwidth@
  \ifdim\dimen@<\tw@\wdz@
   \rlap{\vbox{\normalbaselines\boxz@\vbox to\lineht@{}}}\else
   \rlap{\boxz@}\fi
  \tabskip\displaywidth@\crcr#1\crcr\black@\totwidth@}}
\expandafter\def\csname align (in \string\gather)\endcsname
  #1\endalign{\vcenter{\align@#1\endalign}}
\Invalid@\endalign
\newif\ifxat@
\def\alignat{\RIfMIfI@\DN@{\onlydmatherr@\alignat}\else
 \DN@{\csname alignat \endcsname}\fi\else
 \DN@{\onlydmatherr@\alignat}\fi\next@}
\newif\ifmeasuring@
\newbox\savealignat@
\expandafter\def\csname alignat \endcsname#1#2\endalignat                   
 {\inany@true\xat@false
 \def\tag{\global\tag@true\count@#1\relax\multiply\count@\tw@
  \xdef\tag@{}\loop\ifnum\count@>\and@\xdef\tag@{&\tag@}\advance\count@\m@ne
  \repeat\tag@}%
 \vspace@\allowdisplaybreak@\displaybreak@\intertext@
 \displ@y@\measuring@true                                                   
 \setbox\savealignat@\hbox{$\m@th\displaystyle\Let@
  \attag@{#1}
  \vbox{\halign{\span\preamble@@\crcr#2\crcr}}$}%
 \measuring@false                                                           
 \Let@\attag@{#1}
 \tabskip\centering@\halign to\displaywidth
  {\span\preamble@@\crcr#2\crcr                                             
  \black@{\wd\savealignat@}}}                                               
\Invalid@\endalignat
\def\xalignat{\RIfMIfI@
 \DN@{\onlydmatherr@\xalignat}\else
 \DN@{\csname xalignat \endcsname}\fi\else
 \DN@{\onlydmatherr@\xalignat}\fi\next@}
\expandafter\def\csname xalignat \endcsname#1#2\endxalignat
 {\inany@true\xat@true
 \def\tag{\global\tag@true\def\tag@{}\count@#1\relax\multiply\count@\tw@
  \loop\ifnum\count@>\and@\xdef\tag@{&\tag@}\advance\count@\m@ne\repeat\tag@}%
 \vspace@\allowdisplaybreak@\displaybreak@\intertext@
 \displ@y@\measuring@true\setbox\savealignat@\hbox{$\m@th\displaystyle\Let@
 \attag@{#1}\vbox{\halign{\span\preamble@@\crcr#2\crcr}}$}%
 \measuring@false\Let@
 \attag@{#1}\tabskip\centering@\halign to\displaywidth
 {\span\preamble@@\crcr#2\crcr\black@{\wd\savealignat@}}}
\def\attag@#1{\let\Maketag@\maketag@\let\TAG@\Tag@                          
 \let\Tag@=0\let\maketag@=0
 \ifmeasuring@\def\llap@##1{\setboxz@h{##1}\hbox to\tw@\wdz@{}}%
  \def\rlap@##1{\setboxz@h{##1}\hbox to\tw@\wdz@{}}\else
  \let\llap@\llap\let\rlap@\rlap\fi                                         
 \toks@{\hfil\strut@$\m@th\displaystyle{\@lign\the\hashtoks@}$\tabskip\z@skip
  \global\advance\and@\@ne&$\m@th\displaystyle{{}\@lign\the\hashtoks@}$\hfil
  \ifxat@\tabskip\centering@\fi\global\advance\and@\@ne}
 \iftagsleft@
  \toks@@{\tabskip\centering@&\Tag@\kern-\displaywidth
   \rlap@{\@lign\maketag@\the\hashtoks@\maketag@}%
   \global\advance\and@\@ne\tabskip\displaywidth}\else
  \toks@@{\tabskip\centering@&\Tag@\llap@{\@lign\maketag@
   \the\hashtoks@\maketag@}\global\advance\and@\@ne\tabskip\z@skip}\fi      
 \atcount@#1\relax\advance\atcount@\m@ne
 \loop\ifnum\atcount@>\z@
 \toks@=\expandafter{\the\toks@&\hfil$\m@th\displaystyle{\@lign
  \the\hashtoks@}$\global\advance\and@\@ne
  \tabskip\z@skip&$\m@th\displaystyle{{}\@lign\the\hashtoks@}$\hfil\ifxat@
  \tabskip\centering@\fi\global\advance\and@\@ne}\advance\atcount@\m@ne
 \repeat                                                                    
 \xdef\preamble@{\the\toks@\the\toks@@}
 \xdef\preamble@@{\preamble@}
 \let\maketag@\Maketag@\let\Tag@\TAG@}                                      
\Invalid@\endxalignat
\def\xxalignat{\RIfMIfI@
 \DN@{\onlydmatherr@\xxalignat}\else\DN@{\csname xxalignat
  \endcsname}\fi\else
 \DN@{\onlydmatherr@\xxalignat}\fi\next@}
\expandafter\def\csname xxalignat \endcsname#1#2\endxxalignat{\inany@true
 \vspace@\allowdisplaybreak@\displaybreak@\intertext@
 \displ@y\setbox\savealignat@\hbox{$\m@th\displaystyle\Let@
 \xxattag@{#1}\vbox{\halign{\span\preamble@@\crcr#2\crcr}}$}%
 \Let@\xxattag@{#1}\tabskip\z@skip\halign to\displaywidth
 {\span\preamble@@\crcr#2\crcr\black@{\wd\savealignat@}}}
\def\xxattag@#1{\toks@{\tabskip\z@skip\hfil\strut@
 $\m@th\displaystyle{\the\hashtoks@}$&%
 $\m@th\displaystyle{{}\the\hashtoks@}$\hfil\tabskip\centering@&}%
 \atcount@#1\relax\advance\atcount@\m@ne\loop\ifnum\atcount@>\z@
 \toks@=\expandafter{\the\toks@&\hfil$\m@th\displaystyle{\the\hashtoks@}$%
  \tabskip\z@skip&$\m@th\displaystyle{{}\the\hashtoks@}$\hfil
  \tabskip\centering@}\advance\atcount@\m@ne\repeat
 \xdef\preamble@{\the\toks@\tabskip\z@skip}\xdef\preamble@@{\preamble@}}
\Invalid@\endxxalignat
\newdimen\gwidth@
\newdimen\gmaxwidth@
\def\gmeasure@#1\endgather{\gwidth@\z@\gmaxwidth@\z@\setbox@ne\vbox{\Let@
 \halign{\setboxz@h{$\m@th\displaystyle{##}$}\global\gwidth@\wdz@
 \ifdim\gwidth@>\gmaxwidth@\global\gmaxwidth@\gwidth@\fi
 &\eat@{##}\crcr#1\crcr}}}
\def\gather{\RIfMIfI@\DN@{\onlydmatherr@\gather}\else
 \ingather@true\inany@true\def\tag{&}%
 \vspace@\allowdisplaybreak@\displaybreak@\intertext@
 \displ@y\Let@
 \iftagsleft@\DN@{\csname gather \endcsname}\else
  \DN@{\csname gather \space\endcsname}\fi\fi
 \else\DN@{\onlydmatherr@\gather}\fi\next@}
\expandafter\def\csname gather \space\endcsname#1\endgather
 {\gmeasure@#1\endgather\tabskip\centering@
 \halign to\displaywidth{\hfil\strut@\setboxz@h{$\m@th\displaystyle{##}$}%
 \global\gwidth@\wdz@\boxz@\hfil&
 \setboxz@h{\strut@{\maketag@##\maketag@}}%
 \dimen@\displaywidth\advance\dimen@-\gwidth@
 \ifdim\dimen@>\tw@\wdz@\llap{\boxz@}\else
 \llap{\vtop{\normalbaselines\null\boxz@}}\fi
 \tabskip\z@skip\crcr#1\crcr\black@\gmaxwidth@}}
\newdimen\glineht@
\expandafter\def\csname gather \endcsname#1\endgather{\gmeasure@#1\endgather
 \ifdim\gmaxwidth@>\displaywidth\let\gdisplaywidth@\gmaxwidth@\else
 \let\gdisplaywidth@\displaywidth\fi\tabskip\centering@\halign to\displaywidth
 {\hfil\strut@\setboxz@h{$\m@th\displaystyle{##}$}%
 \global\gwidth@\wdz@\global\glineht@\ht\z@\boxz@\hfil&\kern-\gdisplaywidth@
 \setboxz@h{\strut@{\maketag@##\maketag@}}%
 \dimen@\displaywidth\advance\dimen@-\gwidth@
 \ifdim\dimen@>\tw@\wdz@\rlap{\boxz@}\else
 \rlap{\vbox{\normalbaselines\boxz@\vbox to\glineht@{}}}\fi
 \tabskip\gdisplaywidth@\crcr#1\crcr\black@\gmaxwidth@}}
\newif\ifctagsplit@
\def\CenteredTagsOnSplits{\global\ctagsplit@true}
\def\TopOrBottomTagsOnSplits{\global\ctagsplit@false}
\TopOrBottomTagsOnSplits
\def\split{\relax\ifinany@\let\next@\insplit@\else
 \ifmmode\ifinner\def\next@{\onlydmatherr@\split}\else
 \let\next@\outsplit@\fi\else
 \def\next@{\onlydmatherr@\split}\fi\fi\next@}
\def\insplit@{\global\setbox\z@\vbox\bgroup\vspace@\Let@\ialign\bgroup
 \hfil\strut@$\m@th\displaystyle{##}$&$\m@th\displaystyle{{}##}$\hfill\crcr}
\def\endsplit{\crcr\egroup\egroup\iftagsleft@\expandafter\lendsplit@\else
 \expandafter\rendsplit@\fi}
\def\rendsplit@{\global\setbox9 \vbox
 {\unvcopy\z@\global\setbox8 \lastbox\unskip}
 \setbox@ne\hbox{\unhcopy8 \unskip\global\setbox\tw@\lastbox
 \unskip\global\setbox\thr@@\lastbox}
 \global\setbox7 \hbox{\unhbox\tw@\unskip}
 \ifinalign@\ifctagsplit@                                                   
  \gdef\split@{\hbox to\wd\thr@@{}&
   \vcenter{\vbox{\moveleft\wd\thr@@\boxz@}}}
 \else\gdef\split@{&\vbox{\moveleft\wd\thr@@\box9}\crcr
  \box\thr@@&\box7}\fi                                                      
 \else                                                                      
  \ifctagsplit@\gdef\split@{\vcenter{\boxz@}}\else
  \gdef\split@{\box9\crcr\hbox{\box\thr@@\box7}}\fi
 \fi
 \split@}                                                                   
\def\lendsplit@{\global\setbox9\vtop{\unvcopy\z@}
 \setbox@ne\vbox{\unvcopy\z@\global\setbox8\lastbox}
 \setbox@ne\hbox{\unhcopy8\unskip\setbox\tw@\lastbox
  \unskip\global\setbox\thr@@\lastbox}
 \ifinalign@\ifctagsplit@                                                   
  \gdef\split@{\hbox to\wd\thr@@{}&
  \vcenter{\vbox{\moveleft\wd\thr@@\box9}}}
  \else                                                                     
  \gdef\split@{\hbox to\wd\thr@@{}&\vbox{\moveleft\wd\thr@@\box9}}\fi
 \else
  \ifctagsplit@\gdef\split@{\vcenter{\box9}}\else
  \gdef\split@{\box9}\fi
 \fi\split@}
\def\outsplit@#1$${\align\insplit@#1\endalign$$}
\newdimen\multlinegap@
\multlinegap@1em
\newdimen\multlinetaggap@
\multlinetaggap@1em
\def\MultlineGap#1{\global\multlinegap@#1\relax}
\def\multlinegap#1{\RIfMIfI@\onlydmatherr@\multlinegap\else
 \multlinegap@#1\relax\fi\else\onlydmatherr@\multlinegap\fi}
\def\nomultlinegap{\multlinegap{\z@}}
\def\multline{\RIfMIfI@
 \DN@{\onlydmatherr@\multline}\else
 \DN@{\multline@}\fi\else
 \DN@{\onlydmatherr@\multline}\fi\next@}
\newif\iftagin@
\def\tagin@#1{\tagin@false\in@\tag{#1}\ifin@\tagin@true\fi}
\def\multline@#1$${\inany@true\vspace@\allowdisplaybreak@\displaybreak@
 \tagin@{#1}\iftagsleft@\DN@{\multline@l#1$$}\else
 \DN@{\multline@r#1$$}\fi\next@}
\newdimen\mwidth@
\def\rmmeasure@#1\endmultline{%
 \def\shoveleft##1{##1}\def\shoveright##1{##1}
 \setbox@ne\vbox{\Let@\halign{\setboxz@h
  {$\m@th\@lign\displaystyle{}##$}\global\mwidth@\wdz@
  \crcr#1\crcr}}}
\newdimen\mlineht@
\newif\ifzerocr@
\newif\ifonecr@
\def\lmmeasure@#1\endmultline{\global\zerocr@true\global\onecr@false
 \everycr{\noalign{\ifonecr@\global\onecr@false\fi
  \ifzerocr@\global\zerocr@false\global\onecr@true\fi}}
  \def\shoveleft##1{##1}\def\shoveright##1{##1}%
 \setbox@ne\vbox{\Let@\halign{\setboxz@h
  {$\m@th\@lign\displaystyle{}##$}\ifonecr@\global\mwidth@\wdz@
  \global\mlineht@\ht\z@\fi\crcr#1\crcr}}}
\newbox\mtagbox@
\newdimen\ltwidth@
\newdimen\rtwidth@
\def\multline@l#1$${\iftagin@\DN@{\lmultline@@#1$$}\else
 \DN@{\setbox\mtagbox@\null\ltwidth@\z@\rtwidth@\z@
  \lmultline@@@#1$$}\fi\next@}
\def\lmultline@@#1\endmultline\tag#2$${%
 \setbox\mtagbox@\hbox{\maketag@#2\maketag@}
 \lmmeasure@#1\endmultline\dimen@\mwidth@\advance\dimen@\wd\mtagbox@
 \advance\dimen@\multlinetaggap@                                            
 \ifdim\dimen@>\displaywidth\ltwidth@\z@\else\ltwidth@\wd\mtagbox@\fi       
 \lmultline@@@#1\endmultline$$}
\def\lmultline@@@{\displ@y
 \def\shoveright##1{##1\hfilneg\hskip\multlinegap@}%
 \def\shoveleft##1{\setboxz@h{$\m@th\displaystyle{}##1$}%
  \setbox@ne\hbox{$\m@th\displaystyle##1$}%
  \hfilneg
  \iftagin@
   \ifdim\ltwidth@>\z@\hskip\ltwidth@\hskip\multlinetaggap@\fi
  \else\hskip\multlinegap@\fi\hskip.5\wd@ne\hskip-.5\wdz@##1}
  \halign\bgroup\Let@\hbox to\displaywidth
   {\strut@$\m@th\displaystyle\hfil{}##\hfil$}\crcr
   \hfilneg                                                                 
   \iftagin@                                                                
    \ifdim\ltwidth@>\z@                                                     
     \box\mtagbox@\hskip\multlinetaggap@                                    
    \else
     \rlap{\vbox{\normalbaselines\hbox{\strut@\box\mtagbox@}%
     \vbox to\mlineht@{}}}\fi                                               
   \else\hskip\multlinegap@\fi}                                             
\def\multline@r#1$${\iftagin@\DN@{\rmultline@@#1$$}\else
 \DN@{\setbox\mtagbox@\null\ltwidth@\z@\rtwidth@\z@
  \rmultline@@@#1$$}\fi\next@}
\def\rmultline@@#1\endmultline\tag#2$${\ltwidth@\z@
 \setbox\mtagbox@\hbox{\maketag@#2\maketag@}%
 \rmmeasure@#1\endmultline\dimen@\mwidth@\advance\dimen@\wd\mtagbox@
 \advance\dimen@\multlinetaggap@
 \ifdim\dimen@>\displaywidth\rtwidth@\z@\else\rtwidth@\wd\mtagbox@\fi
 \rmultline@@@#1\endmultline$$}
\def\rmultline@@@{\displ@y
 \def\shoveright##1{##1\hfilneg\iftagin@\ifdim\rtwidth@>\z@
  \hskip\rtwidth@\hskip\multlinetaggap@\fi\else\hskip\multlinegap@\fi}%
 \def\shoveleft##1{\setboxz@h{$\m@th\displaystyle{}##1$}%
  \setbox@ne\hbox{$\m@th\displaystyle##1$}%
  \hfilneg\hskip\multlinegap@\hskip.5\wd@ne\hskip-.5\wdz@##1}%
 \halign\bgroup\Let@\hbox to\displaywidth
  {\strut@$\m@th\displaystyle\hfil{}##\hfil$}\crcr
 \hfilneg\hskip\multlinegap@}
\def\endmultline{\iftagsleft@\expandafter\lendmultline@\else
 \expandafter\rendmultline@\fi}
\def\lendmultline@{\hfilneg\hskip\multlinegap@\crcr\egroup}
\def\rendmultline@{\iftagin@                                                
 \ifdim\rtwidth@>\z@                                                        
  \hskip\multlinetaggap@\box\mtagbox@                                       
 \else\llap{\vtop{\normalbaselines\null\hbox{\strut@\box\mtagbox@}}}\fi     
 \else\hskip\multlinegap@\fi                                                
 \hfilneg\crcr\egroup}
\def\bmod{\mskip-\medmuskip\mkern5mu\mathbin{\fam\z@ mod}\penalty900
 \mkern5mu\mskip-\medmuskip}
\def\pmod#1{\allowbreak\ifinner\mkern8mu\else\mkern18mu\fi
 ({\fam\z@ mod}\,\,#1)}
\def\pod#1{\allowbreak\ifinner\mkern8mu\else\mkern18mu\fi(#1)}
\def\mod#1{\allowbreak\ifinner\mkern12mu\else\mkern18mu\fi{\fam\z@ mod}\,\,#1}
\message{continued fractions,}
\newcount\cfraccount@
\def\cfrac{\bgroup\bgroup\advance\cfraccount@\@ne\strut
 \iffalse{\fi\def\\{\over\displaystyle}\iffalse}\fi}
\def\lcfrac{\bgroup\bgroup\advance\cfraccount@\@ne\strut
 \iffalse{\fi\def\\{\hfill\over\displaystyle}\iffalse}\fi}
\def\rcfrac{\bgroup\bgroup\advance\cfraccount@\@ne\strut\hfill
 \iffalse{\fi\def\\{\over\displaystyle}\iffalse}\fi}
\def\gloop@#1\repeat{\gdef\body{#1}\iterate}
\def\endcfrac{\gloop@\ifnum\cfraccount@>\z@\global\advance\cfraccount@\m@ne
 \egroup\hskip-\nulldelimiterspace\egroup\repeat}
\message{compound symbols,}
\def\binrel@#1{\setboxz@h{\thinmuskip0mu
  \medmuskip\m@ne mu\thickmuskip\@ne mu$#1\m@th$}%
 \setbox@ne\hbox{\thinmuskip0mu\medmuskip\m@ne mu\thickmuskip
  \@ne mu${}#1{}\m@th$}%
 \setbox\tw@\hbox{\hskip\wd@ne\hskip-\wdz@}}
\def\overset#1\to#2{\binrel@{#2}\ifdim\wd\tw@<\z@
 \mathbin{\mathop{\kern\z@#2}\limits^{#1}}\else\ifdim\wd\tw@>\z@
 \mathrel{\mathop{\kern\z@#2}\limits^{#1}}\else
 {\mathop{\kern\z@#2}\limits^{#1}}{}\fi\fi}
\def\underset#1\to#2{\binrel@{#2}\ifdim\wd\tw@<\z@
 \mathbin{\mathop{\kern\z@#2}\limits_{#1}}\else\ifdim\wd\tw@>\z@
 \mathrel{\mathop{\kern\z@#2}\limits_{#1}}\else
 {\mathop{\kern\z@#2}\limits_{#1}}{}\fi\fi}
\def\oversetbrace#1\to#2{\overbrace{#2}^{#1}}
\def\undersetbrace#1\to#2{\underbrace{#2}_{#1}}
\def\sideset#1\and#2\to#3{%
 \setbox@ne\hbox{$\dsize{\vphantom{#3}}#1{#3}\m@th$}%
 \setbox\tw@\hbox{$\dsize{#3}#2\m@th$}%
 \hskip\wd@ne\hskip-\wd\tw@\mathop{\hskip\wd\tw@\hskip-\wd@ne
  {\vphantom{#3}}#1{#3}#2}}
\def\rightarrowfill@#1{\setboxz@h{$#1-\m@th$}\ht\z@\z@
  $#1\m@th\copy\z@\mkern-6mu\cleaders
  \hbox{$#1\mkern-2mu\box\z@\mkern-2mu$}\hfill
  \mkern-6mu\mathord\rightarrow$}
\def\leftarrowfill@#1{\setboxz@h{$#1-\m@th$}\ht\z@\z@
  $#1\m@th\mathord\leftarrow\mkern-6mu\cleaders
  \hbox{$#1\mkern-2mu\copy\z@\mkern-2mu$}\hfill
  \mkern-6mu\box\z@$}
\def\leftrightarrowfill@#1{\setboxz@h{$#1-\m@th$}\ht\z@\z@
  $#1\m@th\mathord\leftarrow\mkern-6mu\cleaders
  \hbox{$#1\mkern-2mu\box\z@\mkern-2mu$}\hfill
  \mkern-6mu\mathord\rightarrow$}
\def\overrightarrow{\mathpalette\overrightarrow@}
\def\overrightarrow@#1#2{\vbox{\ialign{##\crcr\rightarrowfill@#1\crcr
 \noalign{\kern-\ex@\nointerlineskip}$\m@th\hfil#1#2\hfil$\crcr}}}

\def\overleftarrow{\mathpalette\overleftarrow@}
\def\overleftarrow@#1#2{\vbox{\ialign{##\crcr\leftarrowfill@#1\crcr
 \noalign{\kern-\ex@\nointerlineskip}$\m@th\hfil#1#2\hfil$\crcr}}}
\def\overleftrightarrow{\mathpalette\overleftrightarrow@}
\def\overleftrightarrow@#1#2{\vbox{\ialign{##\crcr\leftrightarrowfill@#1\crcr
 \noalign{\kern-\ex@\nointerlineskip}$\m@th\hfil#1#2\hfil$\crcr}}}
\def\underrightarrow{\mathpalette\underrightarrow@}
\def\underrightarrow@#1#2{\vtop{\ialign{##\crcr$\m@th\hfil#1#2\hfil$\crcr
 \noalign{\nointerlineskip}\rightarrowfill@#1\crcr}}}

\def\underleftarrow{\mathpalette\underleftarrow@}
\def\underleftarrow@#1#2{\vtop{\ialign{##\crcr$\m@th\hfil#1#2\hfil$\crcr
 \noalign{\nointerlineskip}\leftarrowfill@#1\crcr}}}
\def\underleftrightarrow{\mathpalette\underleftrightarrow@}
\def\underleftrightarrow@#1#2{\vtop{\ialign{##\crcr$\m@th\hfil#1#2\hfil$\crcr
 \noalign{\nointerlineskip}\leftrightarrowfill@#1\crcr}}}
\message{various kinds of dots,}
\let\DOTSI\relax
\let\DOTSB\relax

\newif\ifmath@
{\uccode`7=`\\ \uccode`8=`m \uccode`9=`a \uccode`0=`t \uccode`!=`h
 \uppercase{\gdef\math@#1#2#3#4#5#6\math@{\global\math@false\ifx 7#1\ifx 8#2%
 \ifx 9#3\ifx 0#4\ifx !#5\xdef\meaning@{#6}\global\math@true\fi\fi\fi\fi\fi}}}
\newif\ifmathch@
{\uccode`7=`c \uccode`8=`h \uccode`9=`\"
 \uppercase{\gdef\mathch@#1#2#3#4#5#6\mathch@{\global\mathch@false
  \ifx 7#1\ifx 8#2\ifx 9#5\global\mathch@true\xdef\meaning@{9#6}\fi\fi\fi}}}
\newcount\classnum@
\def\getmathch@#1.#2\getmathch@{\classnum@#1 \divide\classnum@4096
 \ifcase\number\classnum@\or\or\gdef\thedots@{\dotsb@}\or
 \gdef\thedots@{\dotsb@}\fi}
\newif\ifmathbin@
{\uccode`4=`b \uccode`5=`i \uccode`6=`n
 \uppercase{\gdef\mathbin@#1#2#3{\relaxnext@
  \DNii@##1\mathbin@{\ifx\space@\next\global\mathbin@true\fi}%
 \global\mathbin@false\DN@##1\mathbin@{}%
 \ifx 4#1\ifx 5#2\ifx 6#3\DN@{\FN@\nextii@}\fi\fi\fi\next@}}}
\newif\ifmathrel@
{\uccode`4=`r \uccode`5=`e \uccode`6=`l
 \uppercase{\gdef\mathrel@#1#2#3{\relaxnext@
  \DNii@##1\mathrel@{\ifx\space@\next\global\mathrel@true\fi}%
 \global\mathrel@false\DN@##1\mathrel@{}%
 \ifx 4#1\ifx 5#2\ifx 6#3\DN@{\FN@\nextii@}\fi\fi\fi\next@}}}
\newif\ifmacro@
{\uccode`5=`m \uccode`6=`a \uccode`7=`c
 \uppercase{\gdef\macro@#1#2#3#4\macro@{\global\macro@false
  \ifx 5#1\ifx 6#2\ifx 7#3\global\macro@true
  \xdef\meaning@{\macro@@#4\macro@@}\fi\fi\fi}}}
\def\macro@@#1->#2\macro@@{#2}
\newif\ifDOTS@
\newcount\DOTSCASE@
{\uccode`6=`\\ \uccode`7=`D \uccode`8=`O \uccode`9=`T \uccode`0=`S
 \uppercase{\gdef\DOTS@#1#2#3#4#5{\global\DOTS@false\DN@##1\DOTS@{}%
  \ifx 6#1\ifx 7#2\ifx 8#3\ifx 9#4\ifx 0#5\let\next@\DOTS@@\fi\fi\fi\fi\fi
  \next@}}}
{\uccode`3=`B \uccode`4=`I \uccode`5=`X
 \uppercase{\gdef\DOTS@@#1{\relaxnext@
  \DNii@##1\DOTS@{\ifx\space@\next\global\DOTS@true\fi}%
  \DN@{\FN@\nextii@}%
  \ifx 3#1\global\DOTSCASE@\z@\else
  \ifx 4#1\global\DOTSCASE@\@ne\else
  \ifx 5#1\global\DOTSCASE@\tw@\else\DN@##1\DOTS@{}%
  \fi\fi\fi\next@}}}
\newif\ifnot@
{\uccode`5=`\\ \uccode`6=`n \uccode`7=`o \uccode`8=`t
 \uppercase{\gdef\not@#1#2#3#4{\relaxnext@
  \DNii@##1\not@{\ifx\space@\next\global\not@true\fi}%
 \global\not@false\DN@##1\not@{}%
 \ifx 5#1\ifx 6#2\ifx 7#3\ifx 8#4\DN@{\FN@\nextii@}\fi\fi\fi
 \fi\next@}}}
\newif\ifkeybin@
\def\keybin@{\keybin@true
 \ifx\next+\else\ifx\next=\else\ifx\next<\else\ifx\next>\else\ifx\next-\else
 \ifx\next*\else\ifx\next:\else\keybin@false\fi\fi\fi\fi\fi\fi\fi}
\def\dots{\RIfM@\expandafter\mdots@\else\expandafter\tdots@\fi}
\def\tdots@{\unskip\relaxnext@
 \DN@{$\m@th\mathinner{\ldotp\ldotp\ldotp}\,
   \ifx\next,\,$\else\ifx\next.\,$\else\ifx\next;\,$\else\ifx\next:\,$\else
   \ifx\next?\,$\else\ifx\next!\,$\else$ \fi\fi\fi\fi\fi\fi}%
 \ \FN@\next@}
\def\mdots@{\FN@\mdots@@}
\def\mdots@@{\gdef\thedots@{\dotso@}
 \ifx\next\boldkey\gdef\thedots@\boldkey{\boldkeydots@}\else                
 \ifx\next\boldsymbol\gdef\thedots@\boldsymbol{\boldsymboldots@}\else       
 \ifx,\next\gdef\thedots@{\dotsc}
 \else\ifx\not\next\gdef\thedots@{\dotsb@}
 \else\keybin@
 \ifkeybin@\gdef\thedots@{\dotsb@}
 \else\xdef\meaning@{\meaning\next..........}\xdef\meaning@@{\meaning@}
  \expandafter\math@\meaning@\math@
  \ifmath@
   \expandafter\mathch@\meaning@\mathch@
   \ifmathch@\expandafter\getmathch@\meaning@\getmathch@\fi                 
  \else\expandafter\macro@\meaning@@\macro@                                 
  \ifmacro@                                                                
   \expandafter\not@\meaning@\not@\ifnot@\gdef\thedots@{\dotsb@}
  \else\expandafter\DOTS@\meaning@\DOTS@
  \ifDOTS@
   \ifcase\number\DOTSCASE@\gdef\thedots@{\dotsb@}%
    \or\gdef\thedots@{\dotsi}\else\fi                                      
  \else\expandafter\math@\meaning@\math@                                   
  \ifmath@\expandafter\mathbin@\meaning@\mathbin@
  \ifmathbin@\gdef\thedots@{\dotsb@}
  \else\expandafter\mathrel@\meaning@\mathrel@
  \ifmathrel@\gdef\thedots@{\dotsb@}
  \fi\fi\fi\fi\fi\fi\fi\fi\fi\fi\fi\fi
 \thedots@}
\def\plainldots@{\mathinner{\ldotp\ldotp\ldotp}}
\def\plaincdots@{\mathinner{\cdotp\cdotp\cdotp}}
\def\dotsi{\!\plaincdots@}
\let\dotsb@\plaincdots@
\newif\ifextra@
\newif\ifrightdelim@
\def\rightdelim@{\global\rightdelim@true                                    
 \ifx\next)\else                                                            
 \ifx\next]\else
 \ifx\next\rbrack\else
 \ifx\next\}\else
 \ifx\next\rbrace\else
 \ifx\next\rangle\else
 \ifx\next\rceil\else
 \ifx\next\rfloor\else
 \ifx\next\rgroup\else
 \ifx\next\rmoustache\else
 \ifx\next\right\else
 \ifx\next\bigr\else
 \ifx\next\biggr\else
 \ifx\next\Bigr\else                                                        
 \ifx\next\Biggr\else\global\rightdelim@false
 \fi\fi\fi\fi\fi\fi\fi\fi\fi\fi\fi\fi\fi\fi\fi}
\def\extra@{%
 \global\extra@false\rightdelim@\ifrightdelim@\global\extra@true            
 \else\ifx\next$\global\extra@true                                          
 \else\xdef\meaning@{\meaning\next..........}
 \expandafter\macro@\meaning@\macro@\ifmacro@                               
 \expandafter\DOTS@\meaning@\DOTS@
 \ifDOTS@
 \ifnum\DOTSCASE@=\tw@\global\extra@true                                    
 \fi\fi\fi\fi\fi}
\newif\ifbold@
\def\dotso@{\relaxnext@
 \ifbold@
  \let\next\delayed@
  \DNii@{\extra@\plainldots@\ifextra@\,\fi}%
 \else
  \DNii@{\DN@{\extra@\plainldots@\ifextra@\,\fi}\FN@\next@}%
 \fi
 \nextii@}
\def\extrap@#1{%
 \ifx\next,\DN@{#1\,}\else
 \ifx\next;\DN@{#1\,}\else
 \ifx\next.\DN@{#1\,}\else\extra@
 \ifextra@\DN@{#1\,}\else
 \let\next@#1\fi\fi\fi\fi\next@}
\def\ldots{\DN@{\extrap@\plainldots@}%
 \FN@\next@}
\def\cdots{\DN@{\extrap@\plaincdots@}%
 \FN@\next@}

\def\dotsc{\relaxnext@
 \DN@{\ifx\next;\plainldots@\,\else
  \ifx\next.\plainldots@\,\else\extra@\plainldots@
  \ifextra@\,\fi\fi\fi}%
 \FN@\next@}
\def\cdot{\mathchar"2201 }

\def\mapsto{\DOTSB\mapstochar\rightarrow}

\message{special superscripts,}
\def\dddot#1{{\mathop{#1}\limits^{\vbox to-1.4\ex@{\kern-\tw@\ex@
 \hbox{\rm...}\vss}}}}
\def\ddddot#1{{\mathop{#1}\limits^{\vbox to-1.4\ex@{\kern-\tw@\ex@
 \hbox{\rm....}\vss}}}}
\def\sphat{^{\mathchoice{}{}%
 {\,\,\botsmash{\hbox{\lower4\ex@\hbox{$\m@th\widehat{\null}$}}}}%
 {\,\botsmash{\hbox{\lower3\ex@\hbox{$\m@th\hat{\null}$}}}}}}

\def\spacute{^{\!\botsmash{\hbox{\lower\@ne ex\hbox{\'{}}}}}}
\def\spgrave{^{\mathchoice{}{}{}{\!}%
 \botsmash{\hbox{\lower\@ne ex\hbox{\`{}}}}}}
\def\spdot{^{\hbox{\raise\ex@\hbox{\rm.}}}}
\def\spddot{^{\hbox{\raise\ex@\hbox{\rm..}}}}
\def\spdddot{^{\hbox{\raise\ex@\hbox{\rm...}}}}
\def\spddddot{^{\hbox{\raise\ex@\hbox{\rm....}}}}
\def\spbreve{^{\!\botsmash{\hbox{\lower4\ex@\hbox{\u{}}}}}}

\message{\string\text,}
\def\textonlyfont@#1#2{\def#1{\RIfM@
 \Err@{Use \string#1\space only in text}\else#2\fi}}
\textonlyfont@\rm\tenrm
\textonlyfont@\it\tenit
\textonlyfont@\sl\tensl
\textonlyfont@\bf\tenbf
\def\oldnos#1{\RIfM@{\mathcode`\,="013B \fam\@ne#1}\else
 \leavevmode\hbox{$\m@th\mathcode`\,="013B \fam\@ne#1$}\fi}
\def\text{\RIfM@\expandafter\text@\else\expandafter\text@@\fi}
\def\text@@#1{\leavevmode\hbox{#1}}
\def\mathhexbox@#1#2#3{\text{$\m@th\mathchar"#1#2#3$}}
\def\dag{{\mathhexbox@279}}
\def\ddag{{\mathhexbox@27A}}
\def\S{{\mathhexbox@278}}
\def\P{{\mathhexbox@27B}}
\newif\iffirstchoice@
\firstchoice@true
\def\text@#1{\mathchoice
 {\hbox{\everymath{\displaystyle}\def\textfonti{\the\textfont\@ne}%
  \def\textfontii{\the\textfont\tw@}\textdef@@ T#1}}
 {\hbox{\firstchoice@false
  \everymath{\textstyle}\def\textfonti{\the\textfont\@ne}%
  \def\textfontii{\the\textfont\tw@}\textdef@@ T#1}}
 {\hbox{\firstchoice@false
  \everymath{\scriptstyle}\def\textfonti{\the\scriptfont\@ne}%
  \def\textfontii{\the\scriptfont\tw@}\textdef@@ S\rm#1}}
 {\hbox{\firstchoice@false
  \everymath{\scriptscriptstyle}\def\textfonti
  {\the\scriptscriptfont\@ne}%
  \def\textfontii{\the\scriptscriptfont\tw@}\textdef@@ s\rm#1}}}
\def\textdef@@#1{\textdef@#1\rm\textdef@#1\bf\textdef@#1\sl\textdef@#1\it}
\def\rmfam{0}
\def\textdef@#1#2{%
 \DN@{\csname\expandafter\eat@\string#2fam\endcsname}%
 \if S#1\edef#2{\the\scriptfont\next@\relax}%
 \else\if s#1\edef#2{\the\scriptscriptfont\next@\relax}%
 \else\edef#2{\the\textfont\next@\relax}\fi\fi}
\scriptfont\itfam\tenit \scriptscriptfont\itfam\tenit
\scriptfont\slfam\tensl \scriptscriptfont\slfam\tensl
\newif\iftopfolded@
\newif\ifbotfolded@
\def\topfoldedtext{\topfolded@true\botfolded@false\foldedtext@}
\def\botfoldedtext{\botfolded@true\topfolded@false\foldedtext@}
\def\foldedtext{\topfolded@false\botfolded@false\foldedtext@}
\Invalid@\foldedwidth
\def\foldedtext@{\relaxnext@
 \DN@{\ifx\next\foldedwidth\let\next@\nextii@\else
  \DN@{\nextii@\foldedwidth{.3\hsize}}\fi\next@}%
 \DNii@\foldedwidth##1##2{\setbox\z@\vbox
  {\normalbaselines\hsize##1\relax
  \tolerance1600 \noindent\ignorespaces##2}\ifbotfolded@\boxz@\else
  \iftopfolded@\vtop{\unvbox\z@}\else\vcenter{\boxz@}\fi\fi}%
 \FN@\next@}
\message{math font commands,}
\def\bold{\RIfM@\expandafter\bold@\else
 \expandafter\nonmatherr@\expandafter\bold\fi}
\def\bold@#1{{\bold@@{#1}}}
\def\bold@@#1{\fam\bffam\relax#1}
\def\slanted{\RIfM@\expandafter\slanted@\else
 \expandafter\nonmatherr@\expandafter\slanted\fi}
\def\slanted@#1{{\slanted@@{#1}}}
\def\slanted@@#1{\fam\slfam\relax#1}
\def\roman{\RIfM@\expandafter\roman@\else
 \expandafter\nonmatherr@\expandafter\roman\fi}
\def\roman@#1{{\roman@@{#1}}}
\def\roman@@#1{\fam\rmfam\relax#1}
\def\italic{\RIfM@\expandafter\italic@\else
 \expandafter\nonmatherr@\expandafter\italic\fi}
\def\italic@#1{{\italic@@{#1}}}
\def\italic@@#1{\fam\itfam\relax#1}
\def\Cal{\RIfM@\expandafter\Cal@\else
 \expandafter\nonmatherr@\expandafter\Cal\fi}
\def\Cal@#1{{\Cal@@{#1}}}
\def\Cal@@#1{\noaccents@\fam\tw@#1}
\mathchardef\Gamma="0000
\mathchardef\Delta="0001
\mathchardef\Theta="0002
\mathchardef\Lambda="0003
\mathchardef\Xi="0004
\mathchardef\Pi="0005
\mathchardef\Sigma="0006
\mathchardef\Upsilon="0007
\mathchardef\Phi="0008
\mathchardef\Psi="0009
\mathchardef\Omega="000A
\mathchardef\varGamma="0100
\mathchardef\varDelta="0101
\mathchardef\varTheta="0102
\mathchardef\varLambda="0103
\mathchardef\varXi="0104
\mathchardef\varPi="0105
\mathchardef\varSigma="0106
\mathchardef\varUpsilon="0107
\mathchardef\varPhi="0108
\mathchardef\varPsi="0109
\mathchardef\varOmega="010A
\let\alloc@@\alloc@
\def\hexnumber@#1{\ifcase#1 0\or 1\or 2\or 3\or 4\or 5\or 6\or 7\or 8\or
 9\or A\or B\or C\or D\or E\or F\fi}
\def\loadmsam{%
 \font@\tenmsa=msam10
 \font@\sevenmsa=msam7
 \font@\fivemsa=msam5
 \alloc@@8\fam\chardef\sixt@@n\msafam
 \textfont\msafam=\tenmsa
 \scriptfont\msafam=\sevenmsa
 \scriptscriptfont\msafam=\fivemsa
 \edef\next{\hexnumber@\msafam}%
 \mathchardef\dabar@"0\next39
 \edef\dashrightarrow{\mathrel{\dabar@\dabar@\mathchar"0\next4B}}%
 \edef\dashleftarrow{\mathrel{\mathchar"0\next4C\dabar@\dabar@}}%
 \let\dasharrow\dashrightarrow
 \edef\ulcorner{\delimiter"4\next70\next70 }%
 \edef\urcorner{\delimiter"5\next71\next71 }%
 \edef\llcorner{\delimiter"4\next78\next78 }%
 \edef\lrcorner{\delimiter"5\next79\next79 }%
 \edef\yen{{\noexpand\mathhexbox@\next55}}%
 \edef\checkmark{{\noexpand\mathhexbox@\next58}}%
 \edef\circledR{{\noexpand\mathhexbox@\next72}}%
 \edef\maltese{{\noexpand\mathhexbox@\next7A}}%
 \global\let\loadmsam\empty}%
\def\loadmsbm{%
 \font@\tenmsb=msbm10 \font@\sevenmsb=msbm7 \font@\fivemsb=msbm5
 \alloc@@8\fam\chardef\sixt@@n\msbfam
 \textfont\msbfam=\tenmsb
 \scriptfont\msbfam=\sevenmsb \scriptscriptfont\msbfam=\fivemsb
 \global\let\loadmsbm\empty
 }
\def\widehat#1{\ifx\undefined\msbfam \DN@{362}%
  \else \setboxz@h{$\m@th#1$}%
    \edef\next@{\ifdim\wdz@>\tw@ em%
        \hexnumber@\msbfam 5B%
      \else 362\fi}\fi
  \mathaccent"0\next@{#1}}
\def\widetilde#1{\ifx\undefined\msbfam \DN@{365}%
  \else \setboxz@h{$\m@th#1$}%
    \edef\next@{\ifdim\wdz@>\tw@ em%
        \hexnumber@\msbfam 5D%
      \else 365\fi}\fi
  \mathaccent"0\next@{#1}}
\message{\string\newsymbol,}
\def\newsymbol#1#2#3#4#5{\define#1{}%
  \count@#2\relax \advance\count@\m@ne 
 \ifcase\count@
   \ifx\undefined\msafam\loadmsam\fi \let\next@\msafam
 \or \ifx\undefined\msbfam\loadmsbm\fi \let\next@\msbfam
 \else  \Err@{\Invalid@@\string\newsymbol}\let\next@\tw@\fi
 \mathchardef#1="#3\hexnumber@\next@#4#5\space}

\def\Bbb{\RIfM@\expandafter\Bbb@\else
 \expandafter\nonmatherr@\expandafter\Bbb\fi}
\def\Bbb@#1{{\Bbb@@{#1}}}
\def\Bbb@@#1{\noaccents@\fam\msbfam\relax#1}
\message{bold Greek and bold symbols,}
\def\loadbold{%
 \font@\tencmmib=cmmib10 \font@\sevencmmib=cmmib7 \font@\fivecmmib=cmmib5
 \skewchar\tencmmib'177 \skewchar\sevencmmib'177 \skewchar\fivecmmib'177
 \alloc@@8\fam\chardef\sixt@@n\cmmibfam
 \textfont\cmmibfam\tencmmib
 \scriptfont\cmmibfam\sevencmmib \scriptscriptfont\cmmibfam\fivecmmib
 \font@\tencmbsy=cmbsy10 \font@\sevencmbsy=cmbsy7 \font@\fivecmbsy=cmbsy5
 \skewchar\tencmbsy'60 \skewchar\sevencmbsy'60 \skewchar\fivecmbsy'60
 \alloc@@8\fam\chardef\sixt@@n\cmbsyfam
 \textfont\cmbsyfam\tencmbsy
 \scriptfont\cmbsyfam\sevencmbsy \scriptscriptfont\cmbsyfam\fivecmbsy
 \let\loadbold\empty
}
\def\boldnotloaded#1{\Err@{\ifcase#1\or First\else Second\fi
       bold symbol font not loaded}}
\def\mathchari@#1#2#3{\ifx\undefined\cmmibfam
    \boldnotloaded@\@ne
  \else\mathchar"#1\hexnumber@\cmmibfam#2#3\space \fi}
\def\mathcharii@#1#2#3{\ifx\undefined\cmbsyfam
    \boldnotloaded\tw@
  \else \mathchar"#1\hexnumber@\cmbsyfam#2#3\space\fi}
\edef\bffam@{\hexnumber@\bffam}
\def\boldkey#1{\ifcat\noexpand#1A%
  \ifx\undefined\cmmibfam \boldnotloaded\@ne
  \else {\fam\cmmibfam#1}\fi
 \else
 \ifx#1!\mathchar"5\bffam@21 \else
 \ifx#1(\mathchar"4\bffam@28 \else\ifx#1)\mathchar"5\bffam@29 \else
 \ifx#1+\mathchar"2\bffam@2B \else\ifx#1:\mathchar"3\bffam@3A \else
 \ifx#1;\mathchar"6\bffam@3B \else\ifx#1=\mathchar"3\bffam@3D \else
 \ifx#1?\mathchar"5\bffam@3F \else\ifx#1[\mathchar"4\bffam@5B \else
 \ifx#1]\mathchar"5\bffam@5D \else
 \ifx#1,\mathchari@63B \else
 \ifx#1-\mathcharii@200 \else
 \ifx#1.\mathchari@03A \else
 \ifx#1/\mathchari@03D \else
 \ifx#1<\mathchari@33C \else
 \ifx#1>\mathchari@33E \else
 \ifx#1*\mathcharii@203 \else
 \ifx#1|\mathcharii@06A \else
 \ifx#10\bold0\else\ifx#11\bold1\else\ifx#12\bold2\else\ifx#13\bold3\else
 \ifx#14\bold4\else\ifx#15\bold5\else\ifx#16\bold6\else\ifx#17\bold7\else
 \ifx#18\bold8\else\ifx#19\bold9\else
  \Err@{\string\boldkey\space can't be used with #1}%
 \fi\fi\fi\fi\fi\fi\fi\fi\fi\fi\fi\fi\fi\fi\fi
 \fi\fi\fi\fi\fi\fi\fi\fi\fi\fi\fi\fi\fi\fi}
\def\boldsymbol#1{%
 \DN@{\Err@{You can't use \string\boldsymbol\space with \string#1}#1}%
 \ifcat\noexpand#1A%
   \let\next@\relax
   \ifx\undefined\cmmibfam \boldnotloaded\@ne
   \else {\fam\cmmibfam#1}\fi
 \else
  \xdef\meaning@{\meaning#1.........}%
  \expandafter\math@\meaning@\math@
  \ifmath@
   \expandafter\mathch@\meaning@\mathch@
   \ifmathch@
    \expandafter\boldsymbol@@\meaning@\boldsymbol@@
   \fi
  \else
   \expandafter\macro@\meaning@\macro@
   \expandafter\delim@\meaning@\delim@
   \ifdelim@
    \expandafter\delim@@\meaning@\delim@@
   \else
    \boldsymbol@{#1}%
   \fi
  \fi
 \fi
 \next@}
\def\mathhexboxii@#1#2{\ifx\undefined\cmbsyfam
    \boldnotloaded\tw@
  \else \mathhexbox@{\hexnumber@\cmbsyfam}{#1}{#2}\fi}
\def\boldsymbol@#1{\let\next@\relax\let\next#1%
 \ifx\next\cdot\mathcharii@201 \else
 \ifx\next\prime{{\null\mathcharii@030 \null}}\else
 \ifx\next\lbrack\mathchar"4\bffam@5B \else
 \ifx\next\rbrack\mathchar"5\bffam@5D \else
 \ifx\next\{\mathcharii@466 \else
 \ifx\next\lbrace\mathcharii@466 \else
 \ifx\next\}\mathcharii@567 \else
 \ifx\next\rbrace\mathcharii@567 \else
 \ifx\next\surd{{\mathcharii@170}}\else
 \ifx\next\S{{\mathhexboxii@78}}\else
 \ifx\next\P{{\mathhexboxii@7B}}\else
 \ifx\next\dag{{\mathhexboxii@79}}\else
 \ifx\next\ddag{{\mathhexboxii@7A}}\else
 \DN@{\Err@{You can't use \string\boldsymbol\space with \string#1}#1}%
 \fi\fi\fi\fi\fi\fi\fi\fi\fi\fi\fi\fi\fi}
\def\boldsymbol@@#1.#2\boldsymbol@@{\classnum@#1 \count@@@\classnum@        
 \divide\classnum@4096 \count@\classnum@                                    
 \multiply\count@4096 \advance\count@@@-\count@ \count@@\count@@@           
 \divide\count@@@\@cclvi \count@\count@@                                    
 \multiply\count@@@\@cclvi \advance\count@@-\count@@@                       
 \divide\count@@@\@cclvi                                                    
 \multiply\classnum@4096 \advance\classnum@\count@@                         
 \ifnum\count@@@=\z@                                                        
  \count@"\bffam@ \multiply\count@\@cclvi
  \advance\classnum@\count@
  \DN@{\mathchar\number\classnum@}%
 \else
  \ifnum\count@@@=\@ne                                                      
   \ifx\undefined\cmmibfam \DN@{\boldnotloaded\@ne}%
   \else \count@\cmmibfam \multiply\count@\@cclvi
     \advance\classnum@\count@
     \DN@{\mathchar\number\classnum@}\fi
  \else
   \ifnum\count@@@=\tw@                                                    
     \ifx\undefined\cmbsyfam
       \DN@{\boldnotloaded\tw@}%
     \else
       \count@\cmbsyfam \multiply\count@\@cclvi
       \advance\classnum@\count@
       \DN@{\mathchar\number\classnum@}%
     \fi
  \fi
 \fi
\fi}
\newif\ifdelim@
\newcount\delimcount@
{\uccode`6=`\\ \uccode`7=`d \uccode`8=`e \uccode`9=`l
 \uppercase{\gdef\delim@#1#2#3#4#5\delim@
  {\delim@false\ifx 6#1\ifx 7#2\ifx 8#3\ifx 9#4\delim@true
   \xdef\meaning@{#5}\fi\fi\fi\fi}}}
\def\delim@@#1"#2#3#4#5#6\delim@@{\if#32%
\let\next@\relax
 \ifx\undefined\cmbsyfam \boldnotloaded\@ne
 \else \mathcharii@#2#4#5\space \fi\fi}
\def\vert{\delimiter"026A30C }
\def\Vert{\delimiter"026B30D }
\let\|\Vert

\def\boldkeydots@#1{\bold@true\let\next=#1\let\delayed@=#1\mdots@@
 \boldkey#1\bold@false}  
\def\boldsymboldots@#1{\bold@true\let\next#1\let\delayed@#1\mdots@@
 \boldsymbol#1\bold@false}
\message{Euler fonts,}

\def\frak{\mathfont@\frak}

\def\loadmathfont#1{%
   \expandafter\font@\csname ten#1\endcsname=#110
   \expandafter\font@\csname seven#1\endcsname=#17
   \expandafter\font@\csname five#1\endcsname=#15
   \edef\next{\noexpand\alloc@@8\fam\chardef\sixt@@n
     \expandafter\noexpand\csname#1fam\endcsname}%
   \next
   \textfont\csname#1fam\endcsname \csname ten#1\endcsname
   \scriptfont\csname#1fam\endcsname \csname seven#1\endcsname
   \scriptscriptfont\csname#1fam\endcsname \csname five#1\endcsname
   \expandafter\def\csname #1\expandafter\endcsname\expandafter{%
      \expandafter\mathfont@\csname#1\endcsname}%
 \expandafter\gdef\csname load#1\endcsname{}%
}
\def\mathfont@#1{\RIfM@\expandafter\mathfont@@\expandafter#1\else
  \expandafter\nonmatherr@\expandafter#1\fi}
\def\mathfont@@#1#2{{\mathfont@@@#1{#2}}}
\def\mathfont@@@#1#2{\noaccents@
   \fam\csname\expandafter\eat@\string#1fam\endcsname
   \relax#2}
\message{math accents,}
\def\accentclass@{7}
\def\noaccents@{\def\accentclass@{0}}
\def\makeacc@#1#2{\def#1{\mathaccent"\accentclass@#2 }}
\makeacc@\hat{05E}
\makeacc@\check{014}
\makeacc@\tilde{07E}
\makeacc@\acute{013}
\makeacc@\grave{012}
\makeacc@\dot{05F}
\makeacc@\ddot{07F}
\makeacc@\breve{015}
\makeacc@\bar{016}

\newcount\skewcharcount@
\newcount\familycount@
\def\theskewchar@{\familycount@\@ne
 \global\skewcharcount@\the\skewchar\textfont\@ne                           
 \ifnum\fam>\m@ne\ifnum\fam<16
  \global\familycount@\the\fam\relax
  \global\skewcharcount@\the\skewchar\textfont\the\fam\relax\fi\fi          
 \ifnum\skewcharcount@>\m@ne
  \ifnum\skewcharcount@<128
  \multiply\familycount@256
  \global\advance\skewcharcount@\familycount@
  \global\advance\skewcharcount@28672
  \mathchar\skewcharcount@\else
  \global\skewcharcount@\m@ne\fi\else
 \global\skewcharcount@\m@ne\fi}                                            
\newcount\pointcount@
\def\getpoints@#1.#2\getpoints@{\pointcount@#1 }
\newdimen\accentdimen@
\newcount\accentmu@
\def\dimentomu@{\multiply\accentdimen@ 100
 \expandafter\getpoints@\the\accentdimen@\getpoints@
 \multiply\pointcount@18
 \divide\pointcount@\@m
 \global\accentmu@\pointcount@}
\def\Makeacc@#1#2{\def#1{\RIfM@\DN@{\mathaccent@
 {"\accentclass@#2 }}\else\DN@{\nonmatherr@{#1}}\fi\next@}}
\def\unbracefonts@{\let\Cal@\Cal@@\let\roman@\roman@@\let\bold@\bold@@
 \let\slanted@\slanted@@}
\def\mathaccent@#1#2{\ifnum\fam=\m@ne\xdef\thefam@{1}\else
 \xdef\thefam@{\the\fam}\fi                                                 
 \accentdimen@\z@                                                           
 \setboxz@h{\unbracefonts@$\m@th\fam\thefam@\relax#2$}
 \ifdim\accentdimen@=\z@\DN@{\mathaccent#1{#2}}
  \setbox@ne\hbox{\unbracefonts@$\m@th\fam\thefam@\relax#2\theskewchar@$}
  \setbox\tw@\hbox{$\m@th\ifnum\skewcharcount@=\m@ne\else
   \mathchar\skewcharcount@\fi$}
  \global\accentdimen@\wd@ne\global\advance\accentdimen@-\wdz@
  \global\advance\accentdimen@-\wd\tw@                                     
  \global\multiply\accentdimen@\tw@
  \dimentomu@\global\advance\accentmu@\@ne                                 
 \else\DN@{{\mathaccent#1{#2\mkern\accentmu@ mu}%
    \mkern-\accentmu@ mu}{}}\fi                                             
 \next@}\Makeacc@\Hat{05E}
\Makeacc@\Check{014}
\Makeacc@\Tilde{07E}
\Makeacc@\Acute{013}
\Makeacc@\Grave{012}
\Makeacc@\Dot{05F}
\Makeacc@\Ddot{07F}
\Makeacc@\Breve{015}
\Makeacc@\Bar{016}
\def\Vec{\RIfM@\DN@{\mathaccent@{"017E }}\else
 \DN@{\nonmatherr@\Vec}\fi\next@}
\def\accentedsymbol#1#2{\csname newbox\expandafter\endcsname
  \csname\expandafter\eat@\string#1@box\endcsname
 \expandafter\setbox\csname\expandafter\eat@
  \string#1@box\endcsname\hbox{$\m@th#2$}\define
  #1{\copy\csname\expandafter\eat@\string#1@box\endcsname{}}}
\message{roots,}
\def\sqrt#1{\radical"270370 {#1}}
\let\underline@\underline
\let\overline@\overline
\def\underline#1{\underline@{#1}}
\def\overline#1{\overline@{#1}}
\Invalid@\leftroot
\Invalid@\uproot
\newcount\uproot@
\newcount\leftroot@
\def\root{\relaxnext@
  \DN@{\ifx\next\uproot\let\next@\nextii@\else
   \ifx\next\leftroot\let\next@\nextiii@\else
   \let\next@\plainroot@\fi\fi\next@}%
  \DNii@\uproot##1{\uproot@##1\relax\FN@\nextiv@}%
  \def\nextiv@{\ifx\next\space@\DN@. {\FN@\nextv@}\else
   \DN@.{\FN@\nextv@}\fi\next@.}%
  \def\nextv@{\ifx\next\leftroot\let\next@\nextvi@\else
   \let\next@\plainroot@\fi\next@}%
  \def\nextvi@\leftroot##1{\leftroot@##1\relax\plainroot@}%
   \def\nextiii@\leftroot##1{\leftroot@##1\relax\FN@\nextvii@}%
  \def\nextvii@{\ifx\next\space@
   \DN@. {\FN@\nextviii@}\else
   \DN@.{\FN@\nextviii@}\fi\next@.}%
  \def\nextviii@{\ifx\next\uproot\let\next@\nextix@\else
   \let\next@\plainroot@\fi\next@}%
  \def\nextix@\uproot##1{\uproot@##1\relax\plainroot@}%
  \bgroup\uproot@\z@\leftroot@\z@\FN@\next@}
\def\plainroot@#1\of#2{\setbox\rootbox\hbox{$\m@th\scriptscriptstyle{#1}$}%
 \mathchoice{\r@@t\displaystyle{#2}}{\r@@t\textstyle{#2}}
 {\r@@t\scriptstyle{#2}}{\r@@t\scriptscriptstyle{#2}}\egroup}
\def\r@@t#1#2{\setboxz@h{$\m@th#1\sqrt{#2}$}%
 \dimen@\ht\z@\advance\dimen@-\dp\z@
 \setbox@ne\hbox{$\m@th#1\mskip\uproot@ mu$}\advance\dimen@ 1.667\wd@ne
 \mkern-\leftroot@ mu\mkern5mu\raise.6\dimen@\copy\rootbox
 \mkern-10mu\mkern\leftroot@ mu\boxz@}
\def\boxed#1{\setboxz@h{$\m@th\displaystyle{#1}$}\dimen@.4\ex@
 \advance\dimen@3\ex@\advance\dimen@\dp\z@
 \hbox{\lower\dimen@\hbox{%
 \vbox{\hrule height.4\ex@
 \hbox{\vrule width.4\ex@\hskip3\ex@\vbox{\vskip3\ex@\boxz@\vskip3\ex@}%
 \hskip3\ex@\vrule width.4\ex@}\hrule height.4\ex@}%
 }}}
\message{commutative diagrams,}
\let\ampersand@\relax
\newdimen\minaw@
\minaw@11.11128\ex@
\newdimen\minCDaw@
\minCDaw@2.5pc
\def\minCDarrowwidth#1{\RIfMIfI@\onlydmatherr@\minCDarrowwidth
 \else\minCDaw@#1\relax\fi\else\onlydmatherr@\minCDarrowwidth\fi}
\newif\ifCD@
\def\CD{\bgroup\vspace@\relax\let\ampersand@&\iffalse}\fi
 \CD@true\vcenter\bgroup\Let@\tabskip\z@skip\baselineskip20\ex@
 \lineskip3\ex@\lineskiplimit3\ex@\halign\bgroup
 &\hfill$\m@th##$\hfill\crcr}
\def\endCD{\crcr\egroup\egroup\egroup}
\newdimen\bigaw@
\atdef@>#1>#2>{\ampersand@                                                  
 \setboxz@h{$\m@th\ssize\;{#1}\;\;$}
 \setbox@ne\hbox{$\m@th\ssize\;{#2}\;\;$}
 \setbox\tw@\hbox{$\m@th#2$}
 \ifCD@\global\bigaw@\minCDaw@\else\global\bigaw@\minaw@\fi                 
 \ifdim\wdz@>\bigaw@\global\bigaw@\wdz@\fi
 \ifdim\wd@ne>\bigaw@\global\bigaw@\wd@ne\fi                                
 \ifCD@\enskip\fi                                                           
 \ifdim\wd\tw@>\z@
  \mathrel{\mathop{\hbox to\bigaw@{\rightarrowfill@\displaystyle}}%
    \limits^{#1}_{#2}}
 \else\mathrel{\mathop{\hbox to\bigaw@{\rightarrowfill@\displaystyle}}%
    \limits^{#1}}\fi                                                        
 \ifCD@\enskip\fi                                                          
 \ampersand@}                                                              
\atdef@<#1<#2<{\ampersand@\setboxz@h{$\m@th\ssize\;\;{#1}\;$}%
 \setbox@ne\hbox{$\m@th\ssize\;\;{#2}\;$}\setbox\tw@\hbox{$\m@th#2$}%
 \ifCD@\global\bigaw@\minCDaw@\else\global\bigaw@\minaw@\fi
 \ifdim\wdz@>\bigaw@\global\bigaw@\wdz@\fi
 \ifdim\wd@ne>\bigaw@\global\bigaw@\wd@ne\fi
 \ifCD@\enskip\fi
 \ifdim\wd\tw@>\z@
  \mathrel{\mathop{\hbox to\bigaw@{\leftarrowfill@\displaystyle}}%
       \limits^{#1}_{#2}}\else
  \mathrel{\mathop{\hbox to\bigaw@{\leftarrowfill@\displaystyle}}%
       \limits^{#1}}\fi
 \ifCD@\enskip\fi\ampersand@}
\begingroup
 \catcode`\~=\active \lccode`\~=`\@
 \lowercase{%
  \global\atdef@)#1)#2){~>#1>#2>}
  \global\atdef@(#1(#2({~<#1<#2<}}
\endgroup
\atdef@ A#1A#2A{\llap{$\m@th\vcenter{\hbox
 {$\ssize#1$}}$}\Big\uparrow\rlap{$\m@th\vcenter{\hbox{$\ssize#2$}}$}&&}
\atdef@ V#1V#2V{\llap{$\m@th\vcenter{\hbox
 {$\ssize#1$}}$}\Big\downarrow\rlap{$\m@th\vcenter{\hbox{$\ssize#2$}}$}&&}
\atdef@={&\enskip\mathrel
 {\vbox{\hrule width\minCDaw@\vskip3\ex@\hrule width
 \minCDaw@}}\enskip&}
\atdef@|{\Big\Vert&&}
\atdef@\vert{\Big\Vert&&}
\def\pretend#1\haswidth#2{\setboxz@h{$\m@th\scriptstyle{#2}$}\hbox
 to\wdz@{\hfill$\m@th\scriptstyle{#1}$\hfill}}
\message{poor man's bold,}
\def\pmb{\RIfM@\expandafter\mathpalette\expandafter\pmb@\else
 \expandafter\pmb@@\fi}
\def\pmb@@#1{\leavevmode\setboxz@h{#1}%
   \dimen@-\wdz@
   \kern-.5\ex@\copy\z@
   \kern\dimen@\kern.25\ex@\raise.4\ex@\copy\z@
   \kern\dimen@\kern.25\ex@\box\z@
}
\def\binrel@@#1{\ifdim\wd2<\z@\mathbin{#1}\else\ifdim\wd\tw@>\z@
 \mathrel{#1}\else{#1}\fi\fi}
\newdimen\pmbraise@
\def\pmb@#1#2{\setbox\thr@@\hbox{$\m@th#1{#2}$}%
 \setbox4\hbox{$\m@th#1\mkern.5mu$}\pmbraise@\wd4\relax
 \binrel@{#2}%
 \dimen@-\wd\thr@@
   \binrel@@{%
   \mkern-.8mu\copy\thr@@
   \kern\dimen@\mkern.4mu\raise\pmbraise@\copy\thr@@
   \kern\dimen@\mkern.4mu\box\thr@@
}}
\def\documentstyle#1{\W@{}\input #1.sty\relax}
\message{syntax check,}
\font\dummyft@=dummy
\fontdimen1 \dummyft@=\z@
\fontdimen2 \dummyft@=\z@
\fontdimen3 \dummyft@=\z@
\fontdimen4 \dummyft@=\z@
\fontdimen5 \dummyft@=\z@
\fontdimen6 \dummyft@=\z@
\fontdimen7 \dummyft@=\z@
\fontdimen8 \dummyft@=\z@
\fontdimen9 \dummyft@=\z@
\fontdimen10 \dummyft@=\z@
\fontdimen11 \dummyft@=\z@
\fontdimen12 \dummyft@=\z@
\fontdimen13 \dummyft@=\z@
\fontdimen14 \dummyft@=\z@
\fontdimen15 \dummyft@=\z@
\fontdimen16 \dummyft@=\z@
\fontdimen17 \dummyft@=\z@
\fontdimen18 \dummyft@=\z@
\fontdimen19 \dummyft@=\z@
\fontdimen20 \dummyft@=\z@
\fontdimen21 \dummyft@=\z@
\fontdimen22 \dummyft@=\z@
\def\fontlist@{\\{\tenrm}\\{\sevenrm}\\{\fiverm}\\{\teni}\\{\seveni}%
 \\{\fivei}\\{\tensy}\\{\sevensy}\\{\fivesy}\\{\tenex}\\{\tenbf}\\{\sevenbf}%
 \\{\fivebf}\\{\tensl}\\{\tenit}}
\def\font@#1=#2 {\rightappend@#1\to\fontlist@\font#1=#2 }
\def\dodummy@{{\def\\##1{\global\let##1\dummyft@}\fontlist@}}
\def\nopages@{\output{\setbox\z@\box\@cclv \deadcycles\z@}%
 \alloc@5\toks\toksdef\@cclvi\output}
\let\galleys\nopages@
\newif\ifsyntax@
\newcount\countxviii@
\def\syntax{\syntax@true\dodummy@\countxviii@\count18
 \loop\ifnum\countxviii@>\m@ne\textfont\countxviii@=\dummyft@
 \scriptfont\countxviii@=\dummyft@\scriptscriptfont\countxviii@=\dummyft@
 \advance\countxviii@\m@ne\repeat                                           
 \dummyft@\tracinglostchars\z@\nopages@\frenchspacing\hbadness\@M}
\def\first@#1#2\end{#1}
\def\printoptions{\W@{Do you want S(yntax check),
  G(alleys) or P(ages)?}%
 \message{Type S, G or P, followed by <return>: }%
 \begingroup 
 \endlinechar\m@ne 
 \read\m@ne to\ans@
 \edef\ans@{\uppercase{\def\noexpand\ans@{%
   \expandafter\first@\ans@ P\end}}}%
 \expandafter\endgroup\ans@
 \if\ans@ P
 \else \if\ans@ S\syntax
 \else \if\ans@ G\galleys
 \else\message{? Unknown option: \ans@; using the `pages' option.}%
 \fi\fi\fi}
\def\alloc@#1#2#3#4#5{\global\advance\count1#1by\@ne
 \ch@ck#1#4#2\allocationnumber=\count1#1
 \global#3#5=\allocationnumber
 \ifalloc@\wlog{\string#5=\string#2\the\allocationnumber}\fi}
\def\document{\def\alloclist@{}\def\fontlist@{}}

\let\footnote\undefined
\let\=\undefined
\let\>\undefined

\catcode`\@=\active
\message{... finished}

\input amssym.def
\def\frac#1#2{{#1\over#2}}

\def\fh{\frak {h}}
\def\fc{\frak {c}}
\def\fe{\frak {e}}
\def\fg{\frak {g}}
\def\fso{\text{\sl so}}
\def\fu{\text{\sl u}}
\def\tensor{\otimes}
\def\tr{\text{\rm tr}}
\def\ket#1{|#1\rangle}
\def\bra#1{\langle#1|}

\def\one{{\mathchoice {1\mskip-4mu {\text {\rm l}}} {
1\mskip-4mu {\text {\rm l}}} {1\mskip-4.5mu {\text {\rm l}}} {1\mskip-5mu
{\text {\rm l}}}}}
\nopagenumbers
%
\topinsert
\vskip 3.2 truecm
\endinsert
\centerline{\bigbfont ENTANGLEMENT AS AN OBSERVER-DEPENDENT CONCEPT:}
\centerline{\bigbfont AN APPLICATION TO QUANTUM PHASE TRANSITIONS}
\vskip 20 truept
\centerline{\bigifont G. Ortiz, R. Somma, H. Barnum, E. Knill, and L.
Viola}
\vskip 8 truept
\centerline{\bigrfont Los Alamos National Laboratory} 
\vskip 2 truept
\centerline{\bigrfont Los Alamos, New Mexico 87545, USA} 
\vskip 0.8 truecm

\centerline{\bf 1.  INTRODUCTION}
\vskip 12 truept

In 1935 Einstein, Podolsky, and Rosen [1] published a
Gedanken-experiment  that still surprises us today [2]. Their main
observation highlighted a key feature of quantum phenomena in which  a
pure state of a composite quantum system may cease to be determined by
the states of its constituent subsystems. It is this  non-classical
behavior that Schr\"odinger later termed {\it entanglement} [3].
Entangled states are joint states of two or more distinguishable
quantum systems that cannot be expressed as a mixture of products of
states of each system. To establish whether or not a  state is
entangled requires knowledge of the correlations between subsystems,
and measurement of one subsystem may affect the results of measurements
on another even when the system is in a pure state, something which is
against classical intuition. In Schr\"odinger's words [3]: ``[The] Best
possible knowledge of a {\it whole} does not necessarily include the
same for its {\it parts}.'' Today we know that entanglement is the
defining resource for quantum communications (e.g., allowing quantum
teleportation protocols),  it provides non-local correlations between
subsystems that admit no local  classical interpretation (quantum
non-locality), and  it is believed to be  an essential ingredient to
understanding and unlocking the power of quantum computation [4].

Until very recently, most studies concentrated on how the entanglement
of a number of {\it distinguishable} quantum subsystems  differs from
classical correlations. Typically these studies focus on two subsystems
(the bipartite or ``Alice and Bob'' setting). A notion of entanglement
emerged from these investigations that we will term {\it conventional}
entanglement. This paper addresses the following question: Do we
have a theoretical understanding of entanglement applicable to a full
variety of physical settings? It is clear that not only the assumption
of distinguishability (which is natural and useful in standard quantum
information processing (QIP)  situations), but also the few-subsystem
scenario, are too narrow to embrace  all possible physical settings. In
particular, the need to go beyond this  subsystem-based framework
becomes manifest when one tries to apply the  conventional concept of
entanglement to the physics of matter,  since the constituents of a
quantum many-body system are  {\it indistinguishable} particles. We
shall discuss here a notion of {\it generalized entanglement} (GE),
which  can be applied to any operator  language (fermions, bosons,
spins, etc.) used to describe a physical system and which includes the
conventional entanglement settings introduced to date in a unified
fashion. This is realized by noticing that entanglement is an {\it
observer-dependent} concept, whose properties are determined by the
expectations of a {\it distinguished set of observables} without
reference to a preferred  subsystem decomposition, i.e., it depends on
the physically relevant point  of view.  This viewpoint depends in turn
upon the relationship between different sets of observables that
determine our ability to control the system of interest.  Indeed, the
extent to which entanglement is present depends on the observables used
to measure a system and describe its states. It is worthwhile noting
that the role of observables and control in distinguishing subsystems
such as qubits and determining the emergence of a preferred tensor
product structure has been stressed before by various authors (see
e.g., [5,6]). However, our present approach is  conceptually new and
more general in that the observable sets need {\it not} be associative
algebras, thus the resulting entanglement notion is capable of going
beyond the identification of factorizations of a system into
subsystems. This represents a most conspicuous advantage
as will be highlighted by the condensed-matter application we will
discuss. 

On the other hand, novel materials display complex phase diagrams as a
result of competing  interactions with complex orderings (states of
matter) which are believed to be influenced by being in proximity to
quantum critical points (high-$T_c$, heavy fermion compounds,
frustrated magnets, etc.). And the truth is that our understanding of
the nature of these transitions is very poor. Since entanglement is a
property inherent in quantum states and is strongly related to quantum
correlations, one would expect that the entanglement present in the
ground state of the system changes substantially at a quantum phase
transition (QPT), i.e., when there is a qualitative change in the
behavior of the correlations between constituents. Can we use concepts
and tools borrowed from QIP to better understand QPTs?  What is the
nature and role of entanglement in a QPT? Can one quantify entanglement
in a physically meaningful way?  What properties distinguish between
broken- and non-broken-symmetry QPTs?  These are some of the questions 
we are going to address by example. (See also Refs. [7-10] for recent
work in similar directions.) 

Before we describe the concept of GE, let us start by recalling the
standard framework for conventional entanglement. Given a quantum
system with state space ${\Cal H}$, it can a priori support
inequivalent tensor product structures; for instance, ${\Cal
H}\simeq{\Cal H}_1\tensor{\Cal H}_2\tensor \cdots \tensor{\Cal H}_N$,
or ($N$ even) ${\Cal H}\simeq{\Cal H}_{12}\tensor{\Cal
H}_{34}\tensor \cdots \tensor{\Cal H}_{(N-1)N}$, where ${\Cal
H}_{ij}\simeq{\Cal H}_i\tensor{\Cal H}_j$. Entanglement of states in
${\Cal H}$ is unambiguously defined only once a {\it preferred
subsystem decomposition} is chosen: ${\Cal H} \simeq \bigotimes_j {\Cal
H}_j$. Relative to the selected multipartite structure, a pure state
$\ket{\Psi}$ is entangled iff $\ket{\Psi}$ induces at least some  mixed
subsystem states. For example: if ${\Cal H}\simeq {\Cal H}_A \otimes
{\Cal H}_B$, where the Hilbert spaces for {\it A}lice and {\it B}ob,
${\Cal H}_{A}, {\Cal H}_{B}$, are two-dimensional
($\ket{\psi}_{A,B}=a_0^{A,B} \ket{0}_{A,B}+ a_1^{A,B} \ket{1}_{A,B}$),
and 
$$
\ket{\Psi}=\frac{\ket{0}_A\otimes\ket{1}_B+\ket{1}_A\otimes\ket{0}_B}{\sqrt{2}}
\ ,\ \text{then} \ \ 
\cases{
\rho_A= \tr_B \ket{\Psi}\bra{\Psi} \cr
\rho_B= \tr_A \ket{\Psi}\bra{\Psi}} \ \ \text{are mixed} \ , \eqno(1)
$$
whereas if
$$
\ket{\Psi}=\ket{0}_A\otimes\ket{1}_B \ ,\ \text{then} \ \ 
\rho_A \text{ and } \rho_B \ \ \text{are pure} \ . \eqno(2)
$$
Thus, in the conventional approach, entangled pure states look {\it
mixed}  to at least some {\it local}  observers, while unentangled
pure  states still look pure. In standard QIP settings the choice of
preferred subsystem is usually not problematic, as there is a notion of
locality in  {\it real space}. Indeed, spatially separated local
parties are often taken as synonymous with distinguishable quantum
subsystems.

However, as mentioned above, the distinguishability assumption is too
narrow for a generic physical setting, and yet many questions remain
open:

\noindent
i) What does entanglement mean for indistinguishable particles?

It has been known since the early days of quantum mechanics that
particles in Nature are either bosons or fermions in the
three-dimensional space. Indistinguishability places an additional
constraint on the space of admissible states which manifests itself in
their symmetry properties. The physically accessible state space is the
symmetric(bosons)/antisymmetric(fermions) subspace of the full tensor
product state space. For example, consider the free electron gas (Fermi
liquid) in the real-space $x$-representation ($k_j$ is a label of
momentum), described by a many-body wavefunction of the form 
$$
\!
\langle x_1,x_2,\cdots | \Psi\rangle \sim \text{Det} \pmatrix{e^{ik_1x_1}
& e^{ik_1 x_2} & \cdots \cr e^{ik_2x_1}
& e^{ik_2 x_2} & \cdots \cr \vdots & \vdots & \cdots} \! , 
\langle x_1,x_2,\cdots | \Psi\rangle=-\langle x_2,x_1,\cdots |
\Psi\rangle . \eqno(3)
$$
Exchange correlations are present but are not, in principle, a usable
resource in the usual QIP sense. (Other recent work on generalizing
entanglement to indistinguishable fermions may be found in [11] and
references therein.)

\noindent
ii) Particle or mode entanglement?

The state $\langle x_1,x_2,\cdots | \Psi\rangle$ given above exhibits
mode entanglement with respect to the set of momentum modes; yet, the
electrons are  non-interacting. A set of modes provides a factorization
into a   distinguishable-subsystem structure, but in principle
inequivalent  factorizations (e.g., position modes and momentum modes)
may occur on the  same footing for a given system.

\noindent
iii) What is the most appropriate operator language?

The algebraic language (fermions, bosons, anyons, gauge fields, etc.)
used to describe an interacting quantum system may be intentionally 
changed by mappings such as the Jordan-Wigner transformation and its
generalizations [12].

The bottom line is that nontrivial quantum statistics, or other
physical  restrictions, 
can make the choice of preferred subsystem problematic. Our generalized
approach can overcome such difficulties, by removing the need for a
subsystem decomposition from the start. Even when a subsystem partition
exists, we would like to emphasize that from the viewpoint of GE,
ordinary separable pure states (i.e., states that can be written as a
tensor product of normalized subsystem states, $\ket{\Psi}_s=
\ket{\psi}_1\tensor\ket{\phi}_2\tensor \cdots \tensor\ket{\varphi}_N$)
are not necessarily equivalent to generalized unentangled states. We
will illustrate this issue by example in section 3.2.

\vskip 20 truept

\centerline{\bf 2.  ENTANGLEMENT AS AN OBSERVER-DEPENDENT CONCEPT}
\vskip 12 truept

Our approach to tackle these issues will be to reformulate entanglement
as a {\it subsystem-independent} property that depends on the
physically relevant point of view.  The formal development of GE  may
be found in Refs. [13,14]. Here we will simply recall the key concepts
and further illustrate them by example.

The notions of Hilbert space and linear maps or operators are central
to the formulation of quantum mechanics. Pure states  are elements of a
Hilbert space and physical observables are associated to Hermitian
(self-adjoint) operators which act on that space, and whose eigenvalues
are selected by a measuring apparatus. The role of linear operators in
quantum mechanics is not restricted to the representation of physical
observables. Non-Hermitian operators are very often used as well, for
instance in the context of describing  non-unitary quantum dynamics. 
Linear operators form a complex vector  space under the sum and the
multiplication by a scalar over the field of complex numbers.  If we
augment this vector space with a bilinear operation, the set forms an
{\it algebra}. Quantum mechanics also requires that this operation be
non-commutative and associative. A {\it  Lie algebra} $\fh$ is a vector
space endowed with a bilinear map $[\cdot,\cdot]$ satisfying
antisymmetry and the Jacobi identity
$$
[x, x] = 0 , \  \ [x, [y, z]] + [y, [z, x]] + [z, [x, y]] = 0 
, \ \ \forall \ x, y, z  \in \fh \ . \eqno(4)
$$
An ideal $I$ in a Lie algebra is a subalgebra fixed by the commutator
map: For $x \in I$ and arbitrary $h \in \fh$,  $[h,x] \in I$. A simple
Lie algebra is one with no proper ideals. $\fh$ is said to be
semisimple iff it is the direct sum of simple Lie algebras. A Cartan
subalgebra (CSA) $\fc$ of a semisimple Lie algebra  $\fh$ is a maximal
commutative subalgebra. A vector space carrying  a representation of
$\fh$ decomposes into orthogonal joint  eigenspaces $V_\lambda$ of the
operators in $\fc$. That is, each $V_\lambda$ consists of the set of
states $\ket{\psi}$ such  that for $x\in\fc$,
$x\ket{\psi}=\lambda(x)\ket{\psi}$. The  label $\lambda$ is therefore a
linear functional on $\fc$,  called the weight of $V_\lambda$.   The
subspace of operators in $\fh$ orthogonal in trace inner  product to
$\fc$ can be organized into orthogonal  {\it raising} ($\fe_\mu$)
and {\it lowering} ($\fe_{-\mu}$) operators, which connect different
weight spaces.  For a fixed CSA, the weights form a convex polytope; a
lowest (or highest) weight is an extremal point of such a polytope, and
the one-dimensional weight-spaces having those weights are known as
{\it lowest-weight states}.  The set of lowest-weight states for all
CSAs is the orbit of any one such state under the Lie group generated
by $\fh$.  These are the group-theoretic generalized coherent states
(GCSs) [15], which may be formally represented as 
$$
\ket{\text{GCS}}= e^{\sum_\mu (\eta_\mu \fe_\mu -
\bar{\eta}_\mu \fe_{-\mu})} \ket{\text{ext}} \ , \  \text{where }
\ket{\text{ext}} \ \text{is an extremal state} \  . \eqno(5)
$$  
The GCSs attain {\it minimum uncertainty} in an appropriate
invariant sense [16], thus providing a natural generalization of the
familiar harmonic-oscillator coherent states. 

Our notion of GE is relative to a distinguished subspace of observables
$\Omega$ of the quantum system. The basic idea is to generalize the
observation that standard entangled pure states are those that look
mixed to local observers. Given a pure  state of a system, we remind
the reader that the associated {\it reduced state}, or $\Omega$-state,
is obtained by only considering the expectation values of the
distinguished observables $\Omega$. In other words, an $\Omega$-state
is a linear and positive functional $\omega$ on the operators of
$\Omega$ induced by a density matrix $\rho$ according to
$\omega(x)=\text{tr}(\rho x)$. The set of reduced states is convex. 
We then have the following

\noindent
$\underline{\text{Definition}}:$

A pure state $\ket{\Psi}$ is generalized unentangled relative to the 
distinguished set of observables $\Omega$, if its reduced state is pure
(extremal), and generalized entangled otherwise. Similarly, a mixed
state is generalized unentangled relative to $\Omega$, if it can be
written as a  convex combination of generalized unentangled pure
states.

When $\Omega$ is an irreducibly represented Lie algebra
$\Omega\simeq\fh$, the set of unentangled pure states is identical to
the set of GCSs. Another, more physically transparent, characterization
of generalized-unentangled states is as the set of states that are
unique ground states of a distinguished observable (e.g., a
Hamiltonian).

\vskip 12 truept

\centerline{\bf 2.1 Measure of generalized entanglement}

One would like to have a way to quantify how entangled a given state
is. In our theory of GE (as well as in conventional entanglement
theory), no single measure is able to uniquely characterize  the
entanglement properties of a state. Indeed, a hierarchy of
characterizations of quantum correlations is necessary. In order to
connect the notions of generalized unentanglement and coherence, a
natural  way is to introduce the quadratic purity $P_{\fh}$ relative to
the  distinguished subspace of observables $\fh$.

\noindent
$\underline{\text{Definition}}:$

Let $\{x_i\}$ be a Hermitian ($x_i^\dagger=x_i$), and
commonly-normalized orthogonal basis for $\fh$ ($\text{tr}(x_ix_j)
\propto \delta_{ij}$). For any $\ket{\Psi} \in {\Cal H}$, the
purity of $\ket{\Psi}$  relative to $\fh$, $\fh$-purity, is
$$
P_{\fh}(\ket{\Psi})=\sum_i |\langle \Psi|x_i | \Psi\rangle|^2 \ .
\eqno(6)
$$ 

This measure has the following geometric meaning: $P_{\fh}(\ket{\Psi})$
is the square-distance from 0 of the projection of
$\ket{\Psi}\bra{\Psi}$ onto $\fh$ according to the usual
trace-inner-product norm. If $\fh$ is a Lie algebra, it is invariant
under group transformations, $\tilde{x}_i= \fg^\dagger x_i \fg$, where
$\fg \in e^{i \fh}$. 

\topinsert
\input psfig.sty
\centerline{\hskip2mm\psfig{figure=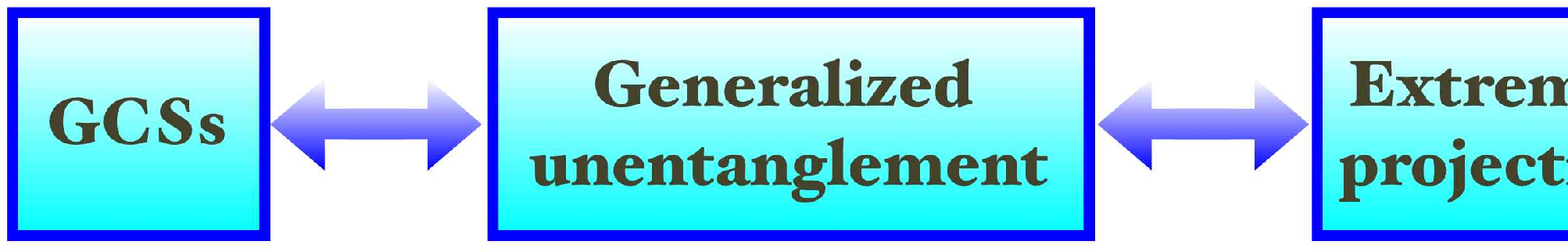,width=14.5truecm,angle=0}}
\noindent
{\bf Figure 1.} {\smallrfont
Equivalent statements in the irreducible Lie-algebraic setting.}
\vskip -2truept
\endinsert

So far, $\fh$ has been assumed to be a {\it real} Lie algebra of
Hermitian operators.  These may be thought of as a preferred family of
Hamiltonians, which generate (via $\fh \mapsto e^{i\fh}$) a Lie group
of unitary operators, or as a preferred family of observables.  

The following characterizations of unentangled states (proven in Ref.
[9]) demonstrate the power of the Lie-algebraic setting (Fig. 1
summarizes equivalent statements for generalized unentangled states).

\noindent
$\underline{\text{Theorem}}:$

The following statements are equivalent for an irreducible representation 
of a Lie algebra $\fh$ on ${\Cal H}$ and state $\rho$

  {\text{\bf (1)}} $\rho$ is generalized unentangled relative to $\fh$.
  
  {\text{\bf (2)}} $\rho=|\Psi\rangle\langle\Psi|$ with 
  $|\Psi\rangle$ the unique ground state of 
  some Hamiltonian $H$ in $\fh$.
  
  {\text{\bf (3)}} $\rho=|\Psi\rangle\langle\Psi|$ with
  $|\Psi\rangle$ a lowest-weight vector of $\fh$.

  {\text{\bf (4)}} $\rho$ has maximum $\fh$-purity.


\vskip 12 truept

\centerline{\bf 3.  A FEW EXAMPLES}
\vskip 12 truept

\centerline{\bf 3.1 Alice and Bob revisited}

As mentioned above, this example corresponds to a bipartite situation
with  preferred distinguishable subsystems $A$ and $B$. The total
Hilbert space ${\Cal H}\simeq{\Cal H}_A\otimes{\Cal H}_B$, such that
$\text{dim }{\Cal H_A} = m$, and $\text{dim }{\Cal H}_B=n$. Physical
access is restricted to {\it local} observables acting on one subsystem
i.e., observables in the set
\vskip -10 truept
$$
\Omega_{\text{\rm loc}}=\{\hat{A}\otimes \one \oplus
 \one\otimes\hat{B}\} \ , \eqno(7)
$$
with $\hat{A}$ and $\hat{B}$ Hermitian and traceless. For each  pure
state $\ket{\Psi} \in {\Cal H}$, the reduced state $\rho_{\text{\rm
red}}$ is then determined by the pair of reduced density operators 
$\rho_A$ and $\rho_B$. 

\noindent
$\underline{\text{Theorem}}$: 

An $\Omega_{\text{\rm loc}}$-state is pure iff it is induced by a pure
product state, meaning that {\it conventional} bipartite entanglement
is equivalent to GE relative to $\Omega_{\text{\rm loc}}$.

For the particular case of two $S$=1/2 parties ($m=n=2$), and
$\Omega=\fh=su(2)\oplus su(2)=\{S_x^j,S_y^j,S_z^j | \ 
j=1,2\}$, where the three generators
$$
S_x^j=\frac{1}{2}\pmatrix{0&1 \cr 1&0}_j ,
S_y^j=\frac{1}{2}\pmatrix{0&-i \cr i&0}_j , 
S_z^j=\frac{1}{2}\pmatrix{1&0 \cr 0&-1}_j \eqno(8)
$$ 
satisfy the $su(2)$ algebra $[S_\alpha^i, S_\beta^j] = i \delta_{ij}
\varepsilon_{\alpha\beta\delta} S_\delta^j \ $
($\varepsilon_{\alpha\beta\delta}$ being the totally antisymmetric
tensor, with $\alpha,\beta,\delta=x,y,z$) and a generic normalized
state
$\ket{\Psi}\!=\!a_1\ket{\!\!\uparrow\uparrow}+a_2\ket{\!\!\uparrow\downarrow}
$ $+a_3\ket{\!\downarrow\uparrow}+a_4 \ket{\!\downarrow\downarrow}$,
the purity relative to $\fh$ is 
\vskip -23 truept
$$
P_{\fh}(\ket{\Psi})=2\sum_{j,\alpha} \langle \Psi |
S_{\alpha}^j|\Psi\rangle^2 \ , \text{ with the result} \cases{P_{\fh}=1
& \text{product states} \cr \vdots & \vdots \cr
P_{\fh}=0 & \text{Bell states}} \ . \eqno(9)
$$

However, when physical access is unrestricted (such as
$\Omega=\fh'=su(4)$), {\it all} pure states have maximal purity meaning
that all of them are generalized unentangled. 

To understand the relevance of the operator language used to describe
our physical system, let us consider a maximally entangled (Bell) state
$$
\ket{\text{\rm
Bell}}=\frac{\ket{\!\uparrow\downarrow}-\ket{\!\downarrow\uparrow}}{\sqrt{2}}=
\frac{S_+^1-S_+^2}{\sqrt{2}} \ket{\!\downarrow\downarrow} \ , \eqno(10)
$$
where $S_\pm^j=S_x^j\pm i S_y^j$, and change our spin into a fermionic
description. After a Jordan-Wigner isomorphic mapping
$$
\cases{
S_+^1=c^\dagger_1 \cr
S_+^2=(1-2n_1) c^\dagger_2} \text{  the Bell state turns into  }
\ket{\text{\rm Bell}}=\frac{c^\dagger_1-c^\dagger_2}{\sqrt{2}} \ket{0}_F
\ , \eqno(11)
$$
where $\ket{0}_F$ is the fermionic vacuum. Written in the fermionic
language this is a single-particle state which can only display {\it
mode} entanglement relative to the local spin algebra $\fh$. However,
had we asked whether $\ket{\text{\rm Bell}}$ is entangled relative to
the algebra
$\fh''=u(2)=\{(c^\dagger_1c^{\;}_2+c^\dagger_2c^{\;}_1)/\sqrt{2},
i(c^\dagger_1c^{\;}_2-c^\dagger_2c^{\;}_1)/\sqrt{2},n_1-\frac{1}{2},n_2
-\frac{1}{2}\}=\{(S^1_xS^2_x+S^1_yS^2_y)/\sqrt{2},(S^1_xS^2_y-S^1_yS^2_x)
/\sqrt{2},S^1_z,S^2_z\}$, the answer would be {\bf No}. The message is 
that the property of entanglement is independent of the operator
language used to describe a state, and only dependent on the observer.

\topinsert
\input psfig.sty
\centerline{\hskip0mm\psfig{figure=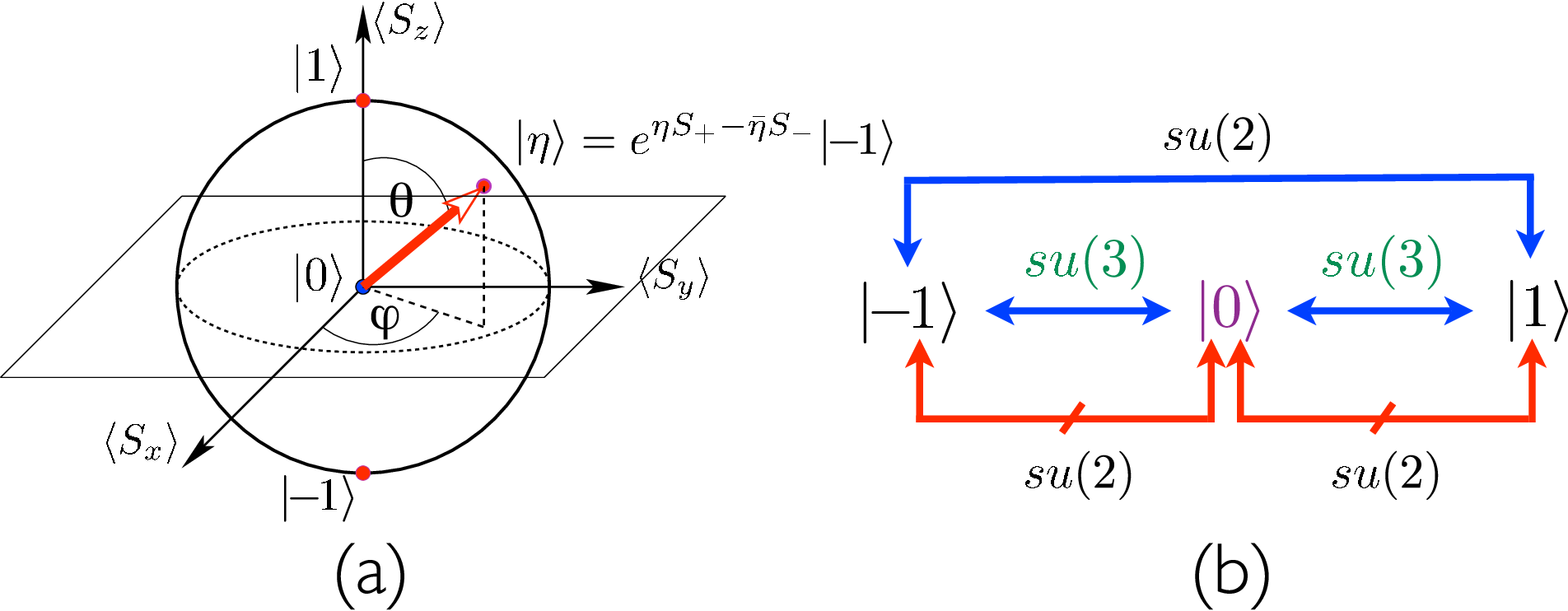,width=15.0truecm,angle=0}}
\noindent
{\bf Figure 2.} (a) {\smallrfont
A ball in $\Bbb{R}^3$ of radius one, whose points (vectors of
expectations values of the $S=1$ operators $S_x, S_y$, and $S_z$)
represent the reduced states. The points on the surface, the extremal
ones, are the {\smallitfont spin-coherent} states $\ket{\eta}$ ($\eta
\in \Bbb{C}$, and $\bar{\eta}$ is its complex conjugate), or the
GCSs of SU(2). These states are generalized unentangled. Notice that
the state $\ket{0}$ is at the center of the ball indicating minimal
distance and, therefore, maximal generalized entanglement.} (b)
{\smallrfont
Schematics of the connections between weight states $\ket{\!-\!1},
\ket{0}, \ket{1}$ by means of group rotations with elements in the
algebras $su(2)$ and $su(3)$. The broken lines indicate that  no group
element exists in SU(2), able to connect either maximal or minimal
weight vectors with $\ket{0}$. On the other hand, there are group
elements in SU(3) connecting them.} 
\vskip -1truept
\endinsert

\vskip 12 truept

\centerline{\bf 3.2 A single spin $S$ = 1}

The extent to which our viewpoint extends the subsystem-based
definition may be appreciated in situations where no physically
natural  subsystem partition exists and conventional entanglement is
not directly applicable. Consider a single spin $S$=1 system, whose
three-dimensional state  space ${\Cal H}$ carries an irreducible
representation of $su(2)$, with generators $S_x, S_y, S_z$. Suppose
that operational access to the system is restricted  to observables in
the given representation of $su(2)$
$$
S_x=\frac{1}{\sqrt{2}}\pmatrix{0&1&0\cr1&0&1\cr0&1&0} ,
S_y=\frac{1}{\sqrt{2}}\pmatrix{0&-i&0\cr i&0&-i\cr0&i&0} ,
S_z=\frac{1}{\sqrt{2}}\pmatrix{1&0&0\cr0&0&0\cr0&0&-1} . \eqno(12)
$$
The reduced states can be identified with vectors of expectation values
of these three observables (see Fig. 2(a)) which form a unit ball in
$\Bbb{R}^3$, $\langle S_x\rangle^2+\langle S_y\rangle^2+\langle
S_z\rangle^2\leq 1$, and the extremal points are those on the surface,
having maximal spin component $1$ for some linear combination of $S_x,
S_y, S_z$.  These are the {\it spin coherent states} $\ket{\eta}$, or
GCSs for SU(2). For any choice of spin direction, ${\Cal H}$ is spanned
by the $|1\rangle, |0\rangle, |{-1}\rangle$ eigenstates of that spin
component; the first and last are GCSs (i.e., unentangled), but
$|0\rangle$ is {\it not}, characterizing $|0\rangle$ as a generalized
({\it maximally}) entangled state relative to $\fh=su(2)$.  Note,
however, that all pure states appear  generalized-unentangled if access
to the full algebra $\fh'=su(3)$ is available  (that is, $su(3)$ is
distinguished).  This is pictorially illustrated in Fig. 2(b).

To appreciate the distinction between separability and generalized
unentanglement, let us consider the case of two spins $S=1$. The only
product states that are generalized unentangled  with respect to
$\fh=su(2)\oplus su(2)$ are those that are products of local SU(2) spin
coherent states, e.g., the state $\ket{1}\tensor \ket{1}$. Instead,
product states such as  $\ket{0}\tensor \ket{0}$ will be entangled (in
particular, this state is maximally entangled).

\vskip 16 truept

\centerline{\bf 4.  QUANTUM PHASE TRANSITIONS AND ENTANGLEMENT}
\vskip 12 truept

QPTs occur at zero temperature due to changes in the interactions among
the constituents of the system. Each quantum phase is characterized by
different behaviors of appropriate correlation functions (such as,
long-range order, algebraic or exponential decay). Since the notions of
entanglement and quantum correlations are intimately related, it
becomes quite natural to look for a good measure of entanglement as a 
way to characterize the QPT.

In this section we study the paramagnetic-to-ferromagnetic QPT present
in the one-dimensional ($S=1/2$) anisotropic XY model in a transverse
magnetic field, applying the concepts of GE described in the previous
sections (a more expanded discussion is also  given in [17]).
Fortunately, this model is exactly solvable and the ground state
properties are well known, allowing for a better understanding of the
entanglement properties at and across the  transition point.

\topinsert
\input psfig.sty
\centerline{\hskip0mm\psfig{figure=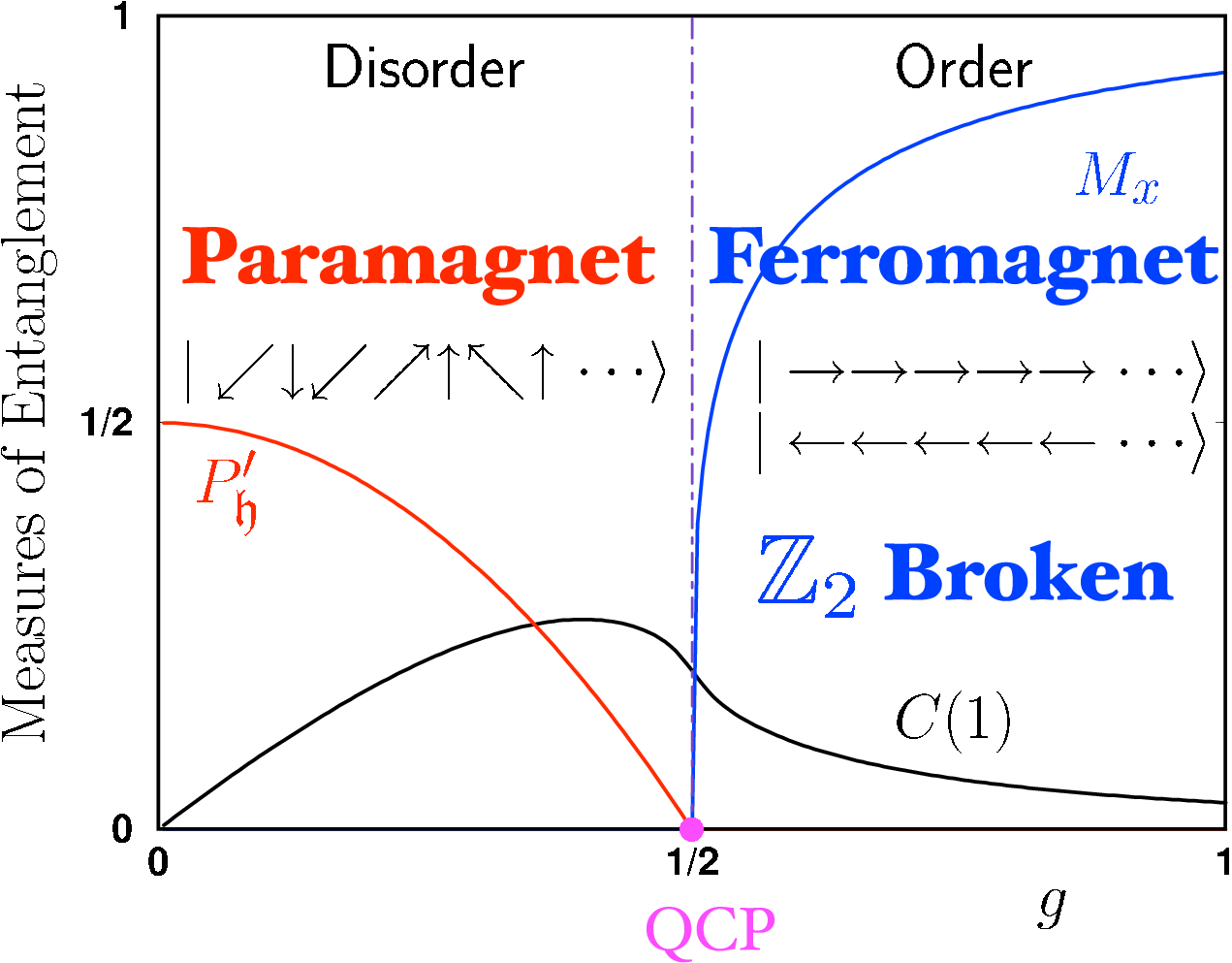,width=8.0truecm,angle=0}}
\noindent
{\bf Figure 3.} {\smallrfont
Measures of entanglement for the Ising model in a transverse magnetic
field ($\gamma=1$). The blue line represents the magnetization in the
$x$-direction $M_x$ (order parameter), the black line is the nearest
neighbors concurrence $C(1)$, and the red line identifies the purity
relative to the $\fu(N)$ Lie algebra, $P'_{\fh}=P_{\fh}-1/(1+\gamma)$. 
QCP stands for quantum critical point.}
\vskip -0truept
\endinsert

The model Hamiltonian for an $N$ ($N$ even) spin system is given
by
\vskip -0.4cm
$$
H= -g \sum\limits_{j=1}^N [(1+\gamma) \sigma_x^j \sigma_x^{j+1} +
(1-\gamma) \sigma_y^j \sigma_y^{j+1} ] + \sum\limits_{j=1}^N \sigma_z^j
\ , \eqno(13)
$$
\vskip -0.3cm \noindent
where the Pauli spin operators $\sigma_\alpha^j = 2 S_\alpha^j$ (see
Eq. (8)), $g \in [0,\infty)$ is the ratio between the spin-spin
exchange coupling and the external magnetic field, and $\gamma \in
[0,1]$ is the anisotropy in the $xy$-plane.  Periodic boundary
conditions will be assumed (i.e.,
$\sigma_\alpha^{N+1}=\sigma_\alpha^1$). The operator $Z_2=\prod_{j=1}^N
\sigma_z^j$ that maps $(\sigma_x^j , \sigma_y^j , \sigma_z^j)
\rightarrow (-\sigma_x^j , -\sigma_y^j , \sigma_z^j) \ ,\forall j \ $, 
is a global ($\Bbb{Z}_2$) symmetry of the model, leaving the
Hamiltonian invariant ($[H,Z_2]=0$). Therefore, for finite $N$, the
non-degenerate
energy eigenstates $\ket{E}$ satisfy $Z_2 \ket{E} = z_2 \ket{E}$, where
$z_2=\pm 1$. The ground state $\ket{G}$ belongs to the sector $z_2=+1$.
For $\gamma=1$ the model reduces to the Ising model in a transverse
magnetic field. In this case and for $g \rightarrow 0$ the spins align
in the direction of the external magnetic field ($z$-direction), so the
ground state is $\ket{G}=\ket {\!\!\downarrow \downarrow \cdots
\downarrow}$. On the other hand, when $g\rightarrow \infty$ the
external field becomes irrelevant and the spins align in the
$x$-direction:  $\ket{G} \simeq \ket {\!\!\rightarrow \rightarrow
\cdots \rightarrow}+\ket {\!\!\leftarrow \leftarrow \cdots
\leftarrow}$, which becomes degenerate  with $\ket{G_-} \simeq \ket
{\!\!\rightarrow \rightarrow \cdots \rightarrow}-\ket {\!\!\leftarrow
\leftarrow \cdots \leftarrow}$ in the thermodynamic limit ($N
\rightarrow \infty$). Therefore, the system undergoes a QPT at some
critical value $g_c$ such that for $g>g_c$ a long range order in the
$x$-direction develops ($\langle \sigma_x^1 \sigma_{x}^{N/2} \rangle
=M_x^2 \ne 0$, where $M_x$ is the total magnetization in the
$x$-direction) and the global $\Bbb{Z}_2$ symmetry breaks
spontaneously. In other words, for $g>g_c$ the quantum phase is
ferromagnetic (see Fig. 3).

The situation is similar for other anisotropies $\gamma$. In principle,
one would expect that the critical point $g_c$ depends on the
anisotropy.  Remarkably, this is not the case and $g_c=1/2$,  $\forall
\gamma$ such that $1 \ge \gamma>0$. (For $\gamma=0$, there is a second
order {\it non-broken symmetry} QPT, i.e. with no symmetry order
parameter, between an {\it insulator} and a quantum critical phase
[17]. The universality class is different than for $\gamma >0$.)

\vskip 12 truept

\centerline{\bf 4.1 The hydrogen atom of QPT}

As mentioned before, the anisotropic XY model in a transverse magnetic
field is exactly solvable. Finding its ground state is a simple task if
we first map the Pauli spin operators into spinless fermionic operators
through the Jordan-Wigner formula
\vskip-0.3cm
$$
S_+^j = \frac{\sigma_x^j + i \sigma_y^j}{2} = \left (
\prod\limits_{k=1}^{j-1} -\sigma_z^k \right ) c^\dagger_j \  , \ 
\sigma_z^j = 2n_j -1 \ , \eqno(14)
$$
where the operators $c^\dagger_j$ and
$c^{\;}_j=(c^\dagger_j)^\dagger$,  create and destroy a fermion at site
$j$, respectively, and $n_j= c^\dagger_j c^{\;}_j$ is the fermionic
number operator. 

Since the model exhibits translational symmetry, it is useful to map
the fermionic operators in the site representation to the
Fourier-transformed operators; i.e., to
\vskip -10 truept
$$
{c}^\dagger_k = \frac{1}{\sqrt{N}} \sum\limits_{j=1}^N e^{-ikj}
c^\dagger_j \ , \eqno(15)
$$
\vskip -12 truept
\noindent
where the operator ${c}^\dagger_k$ 
(${c}^{\;}_k=({c}^\dagger_k)^\dagger$) creates (destroys) a fermion in
the mode $k$. In the fermionic language the sector $z_2=+1$ corresponds
to fermionic anti-periodic boundary conditions and thus $k \in K= K_+ +
K_- =  \left [ \pm \frac{\pi}{N},
\pm\frac{3\pi}{N},\cdots,\pm\frac{(N-1)\pi}{N} \right ]$.  Since the
ground state  $\ket{G}$ belongs to this sector we will only consider
these $2^{N-1}$ states. In this new operator algebra, the Hamiltonian
possess terms that correlate the $k$ and $-k$ modes (Cooper pairs). At
this point, the Hamiltonian can be diagonalized by performing a
Bogoliubov transformation; that is, by defining a new set of fermionic
(quasiparticle) operators $\gamma^\dagger_k = u_k {c}^\dagger_k + i v_k
{c}^{\;}_{-k}$,  with $u_k^2 + v_k^2 =1$, $u_k=u_{-k}$, and
$v_k=-v_{-k}$. The transformed  Hamiltonian reads
\vskip -10 truept
$$
H = \sum\limits_{k \in K} \epsilon_k (\gamma^\dagger_k \gamma^{\;}_k
-1/2) \ , \eqno(16)
$$
\vskip -12 truept
\noindent
where $\gamma^\dagger_k \gamma^{\;}_k$ is the quasiparticle number
operator. Since the energy per mode $\epsilon_k = 2 \sqrt{(-1+2 g
\cos k)^2+ 4 g^2 \gamma^2 \sin^2 k} > 0$  $\forall k \in K$,  the
ground state of the system contains no quasiparticles: $\gamma^{\;}_k
\ket{G} =0$. Thus, $\ket{G}=\ket{\text{BCS}}= \prod\limits_{k \in
K_+} (u_k + i v_k c^\dagger_k c^\dagger_{-k}) \ket{0}_F$,  where
$\ket{0}_F$ is the fermionic vacuum state (i.e., $c^{\;}_k
\ket{0}_F=0$), and the name $\ket{\text{BCS}}$ refers to the
Bardeen-Cooper-Schrieffer superconducting state.

\vskip 12 truept

\centerline{\bf 4.2 $\fh$-Purity as an indicator of the QPT}

Figure 3 displays the behavior of the concurrence $C(1)$ (a measure of
bipartite  entanglement) between nearest neighbors for the case
$\gamma=1$ (Ising model  in a transverse magnetic field) as a function
of $g$ (see also [8]). Although  it is possible to show that its
derivative, $\partial_g C(1)$, diverges at the critical point 
($g_c=1/2$), the function itself has a smooth behavior across $g_c$,
thus it does not directly characterize the QPT.  

The relevant question is whether GE may be able to characterize the
QPT, and to what extent. In order to answer this question  we first
have to identify the relevant algebra $\fh$ of observables that should
be considered. Since in terms of fermionic operators (after the
Jordan-Wigner transformation) the Hamiltonian of the system is a linear
combination of quadratic operators belonging to the Lie algebra 
$\fso(2N) = \{ c^\dagger_k c^{\;}_{k'}, c^\dagger_k c^\dagger_{k'},
c^{\;}_k c^{\;}_{k'}\space \space | \ \forall \ k,k' \in K \}$, the
ground state (if not degenerate) must be a GCS of this algebra. In
other words, the $\ket{\text{BCS}}$ state can be obtained by applying
$\fso (2N)$ group operations to the fully polarized state (which is
also a GCS  of $\fso (2N)$): $\ket{\text{BCS}} =  \prod\limits_{k \in
K_+} e^{i \phi_k c^\dagger_k c^\dagger_{-k} }  \ket{0}_F$, with $u_k =
\cos \phi_k$ and $v_k = \sin \phi_k$. In this way, the $\fh$-purity
relative to this algebra is always maximum ($P_{\fh}=1$, $\forall \ 
g,\gamma$)  and does not give information about the QPT.

The $\ket{\text{BCS}}$ state has no longer the property of being GCS 
and becomes generalized entangled ($P_{\fh} \le 1$) when we choose a  Lie
subalgebra of $\fso(2N)$ that does not contain operators of the form
$c^\dagger_k c^\dagger_{-k}$.  This is the case if we consider the 
subalgebra that conserves the number of fermions; that is
$\fu(N)=\{c^\dagger_k c^{\;}_{k'} \space | \ \forall k,k' \in K \}$.
In particular, in the $\ket{\text{BCS}}$ state the expectation values
of the operators in $\fu(N)$ are $\langle c^\dagger_k c^{\;}_{k'}
\rangle = \delta_ {kk'} v_k^2$. The $\fh$-purity becomes 
\vskip -10 truept
$$
P_{\fh}(\ket{\text{BCS}}) = \frac{4}{N} \sum
\limits_{k \in K} \langle c^\dagger_k c^{\;}_{k} -1/2 \rangle^2
=\frac{4}{N}\sum \limits_{k \in K} (v_k^2 -1/2)^2 \ , \eqno(17)
$$
where the constant 1/2 has been subtracted for normalization reasons. 
Then, any fermionic product state of the form $\ket{\psi} =
c^\dagger_{k_1}c^\dagger_{k_2} \cdots c^\dagger_{k_m}\ket{0}_F$ is a
GCS of this algebra, thus, it has maximum purity ($P_{\fh}=1$).

The $\fh$-purity relative to the $\fu(N)$ algebra for the
$\ket{\text{BCS}}$ state can be exactly calculated in the thermodynamic
limit (where the sum becomes an integral), yielding 
\vskip -10 truept
$$
P_{\fh}(\ket{\text{BCS}}) =\cases{ 
\frac{1}{1-\gamma^2} \left[ 1- \frac{\gamma^2} {\sqrt{
1-4g^2(1-\gamma^2)}} \right] & \text{for $g \le g_c$} \cr
\frac{1}{1+\gamma} & \text{for $g>g_c$}} \ . \eqno(18)
$$
Obviously, $P_{\fh}=1$ for $g=0$, where the state $\ket{\text{BCS}}$
reduces to  $\ket{0}_F \equiv \ket{\!\!\downarrow \downarrow \cdots
\downarrow}$. On the other hand, the $\fh$-purity can be related with
the fluctuations in the total number of fermions in this model, 
\vskip -10 truept
$$
P_{\fh}(\ket{\text{BCS}})=1-\frac{2}{N}(\langle  \hat{N}^2 \rangle -
\langle \hat{N} \rangle^2) \ , \ \text{where }\hat{N}=\sum_{k
\in K}c^\dagger_k c^{\;}_k  \ .\eqno(19)
$$
\vskip -9 truept

Figure 4(a) shows the (shifted) $\fh$-purity vs $g$, for different
anisotropies $\gamma$. Remarkably, this function behaves as a {\it
disorder parameter}, being 0 in the ferromagnetic (ordered)  phase and
different from 0 in the paramagnetic (disordered) phase, sharply
identifying the QPT. (For $\gamma=0$, the $\fh$-purity with respect to
$\bigoplus_j su(2)_j$ is a good indicator of the non-broken symmetry
QPT [17].)

The critical behavior of the $\fh$-purity can also be extracted by 
performing a Taylor-expansion near the critical point (in the
disordered phase, $\gamma>0$). One finds $P_{\fh} \sim (g_c-g)^\nu$ (Fig. 4(b)), 
with $\nu=1$ being the critical exponent that characterizes the Ising
universality class; that is, the correlation length  $\ell$ in this
model diverges with the same exponent (Fig. 4(b)):  $\ell \sim
(g_c-g)^{-\nu}$.

All these interesting features make the quadratic purity measure of GE
a physically appealing measure that characterizes the QPT. In
principle, for other kinds of systems one could think  in terms of
different algebras of observables. However, identifying the appropriate
algebra remains in general an open problem.
\topinsert
\input psfig.sty
\centerline{\hskip0mm\psfig{figure=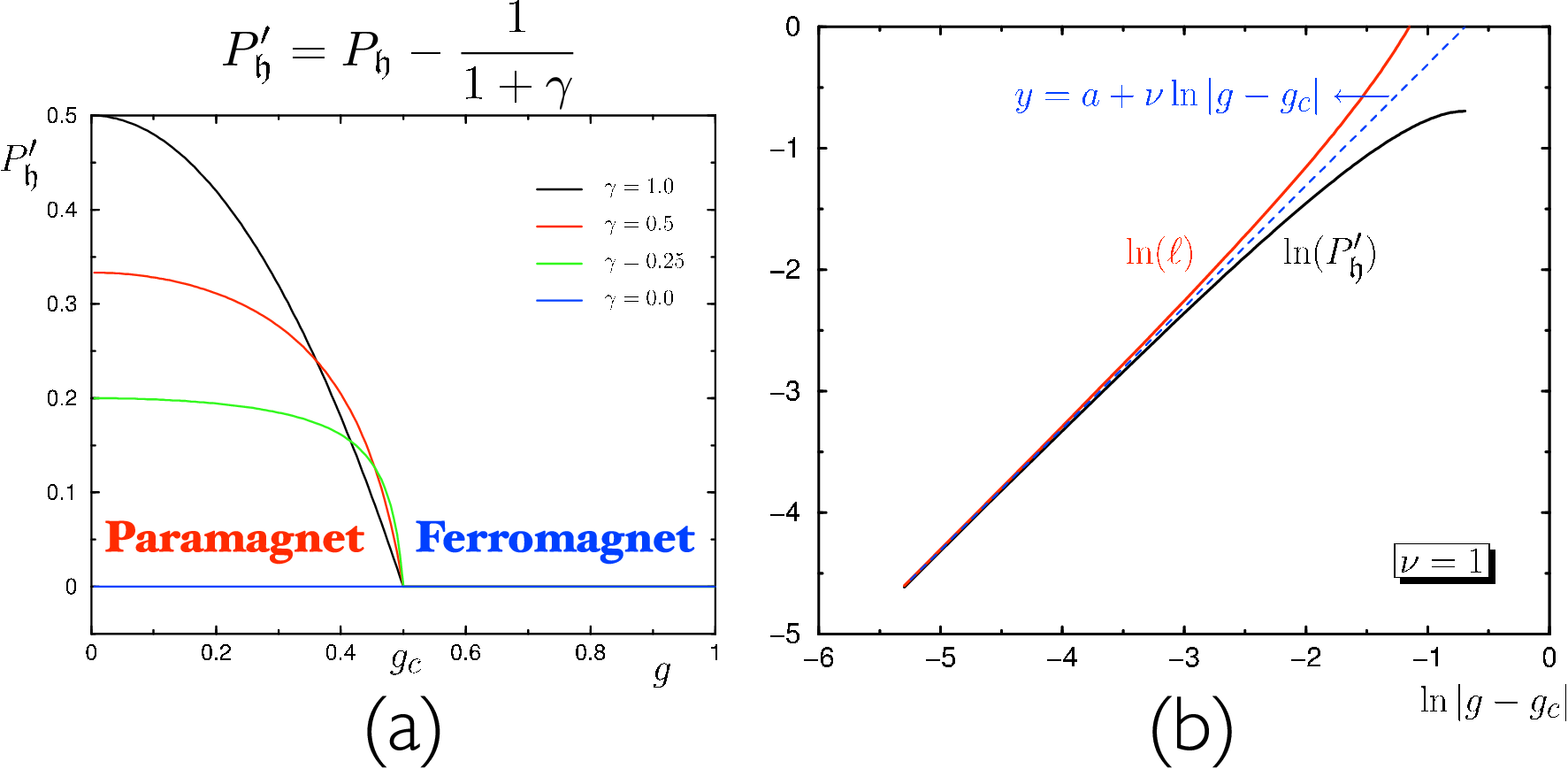,width=15.0truecm,angle=0}}
\noindent
{\bf Figure 4.} (a) {\smallrfont 
Purity (shifted) relative to the $\fu(N)$ algebra in the ground state
$\ket{\text{BCS}}$ of the anisotropic XY model in a transverse
magnetic field.} (b) {\smallrfont Scaling behavior of the purity
relative to the $\fu(N)$ algebra.}
\vskip -12truept
\endinsert

\vskip 18 truept

\centerline{\bf 4.  CONCLUSIONS}
\vskip 8 truept

Entanglement is of one of the key features of quantum mechanics, and 
consequently a deep understanding of its properties and physical 
implications is essential not only to fundamentally advance our 
understanding of how Nature works, but also to design more powerful 
technologies.  In this work we have described a generalization of 
conventional entanglement which is applicable to arbitrary physical 
settings, going beyond the standard distinguishable-subsystem approach. 
It is clear from our formulation that entanglement is a (relative) 
context-dependent notion, whose properties are determined by the 
expectations of a distinguished set of observables $\Omega$ of the 
system of interest, thus being observer-dependent.  The concept is 
independent of the operator language used to describe the system but 
depends upon the way we can physically access information. (In the 
QIP setting locality is usually involved.) Most importantly, we
have tied together the theory of entanglement and the theory of
generalized coherent states in those cases where the set $\Omega$ 
forms a Lie algebra $\fh$, thus establishing that generalized
unentangled states are the (highest-weight) generalized coherent
states of $\fh$. We have also introduced measures of generalized
entanglement. In particular, we introduced the purity relative to a
subalgebra which can represent a global measure of entanglement.
Different choices of observable subalgebras may provide different
information on the hierarchical structure of quantum correlations. 

In the second part of the paper we discussed an application of our
theory of generalized entanglement to the study of quantum phase
transitions. We showed that the purity relative to a subalgebra can be
used as a quantum phase transition indicator. Indeed, it is a useful
diagnostic tool playing the role of a disorder  (or order [7]) 
parameter for broken-symmetry quantum phase transitions. Several 
important open questions remain.  In particular:  Given the model
Hamiltonian describing a physical system, what is the minimum number of
observable algebras $\fh$ one needs to fully characterize a  quantum
critical point in arbitrary space dimensions? How would one choose
them?

The research described herein was sponsored by the US DOE. GO is
grateful to the US Army Research Office for travel support. 

\vskip 12 truept

\centerline{\bf REFERENCES}
\vskip 10 truept

\item{[1]}
A. Einstein, B. Podolsky, and N. Rosen, Phys. Rev. {\bf 47}, 777 (1935).
\item{[2]}
J. S. Bell, {\it Speakable and Unspeakable in Quantum Mechanics},
(Cambridge Univ. Press, Cambridge, 1993). 
\item{[3]}
E. Schr\"odinger, Naturwissenschaften {\bf 23}, 807 (1935), {\it ibid}
823 (1935), {\it ibid} 844 (1935).
\item{[4]}
M. A. Nielsen and I. L. Chuang, {\it Quantum Computation and Quantum
Information} (Cambridge Univ. Press, Cambridge, 2000).
\item{[5]}
L. Viola, E. Knill, and R. Laflamme, J. Phys. A: Math. Gen {\bf 34},
7067 (2001).
\item{[6]}
P. Zanardi, Phys. Rev. Lett. {\bf 87}, 077901 (2001).
\item{[7]}
T. J. Osborne and M. A. Nielsen, Quant. Inf. Process. {\bf 1}, 45 (2002);
Phys. Rev. A {\bf 66}, 032110 (2002).
\item{[8]}
A. Osterloh, L. Amico, G. Falci, and R. Fazio, Nature {\bf 416}, 608 (2002).
\item{[9]}
M. Arnesen, S. Bose, and V. Vedral, Phys. Rev. Lett. {\bf 87}, 290 (2001).
\item{[10]}
G. Vidal, J. I. Latorre, E. Rico, and A. Kitaev, Phys. Rev. Lett. {\bf 90},
227902 (2003).
\item{[11]}
K. Eckert, J. Schliemann, D. Bru{\ss}, and M. Lewenstein, Ann. Phys.
{\bf 299}, 88 (2002).
\item{[12]}
C. D. Batista and G. Ortiz, Phys. Rev. Lett. {\bf 86}, 1082 (2001);
cond-mat/0207106, Adv. Phys. {\bf 53}, 1 (2004); G. Ortiz and C. D. 
Batista in {\it Recent Progress in Many-Body Theories}, Eds. R. F. Bishop, 
T. Brandes, K. A. Gernoth, N. R. Walet, and Y. Xian (World Scientific,
Singapore, 2002), p. 425.
\item{[13]}
H. Barnum, E. Knill, G. Ortiz, and L. Viola, Phys. Rev. A {\bf 68},
032308 (2003).
\item{[14]}
H. Barnum, E. Knill, G. Ortiz, R. Somma, and L. Viola, quant-ph/0305023,
Phys. Rev. Lett. {\bf 92}, 107902 (2004). 
\item{[15]}
A. Perelomov, {\it Generalized Coherent States and their Applications},
(Springer Verlag, Berlin, 1986).
\item{[16]}
R. Delbourgo and J. R. Fox, J. Phys. A, {\bf 10} L233 (1977); 
R. Delbourgo, J. Phys. A {\bf 10}, 1837 (1977).
\item{[17]}
R. Somma, G. Ortiz, H. Barnum, E. Knill, and L. Viola,
quant-ph/0403035. 

\end